# Transcribing RNA polymerases :
# Dynamics of Twin Supercoiled Domains

Marc JOYEUX

*Laboratoire Interdisciplinaire de Physique*

*CNRS and Université Grenoble Alpes*

*38400 St Martin d'Hères*

*France*

email : marc.joyeux@univ-grenoble-alpes.fr

**Running title** : Dynamics of Twin Supercoiled Domains

**ABSTRACT** : Gene transcription by a RNA Polymerase (RNAP) enzyme requires that double-stranded DNA be locally and transiently opened, which results in an increase of DNA supercoiling downstream of the RNAP and a decrease of supercoiling upstream of it. When the DNA is initially torsionally relaxed and the RNAP experiences sufficiently large rotational drag, these variations lead to positively supercoiled plectonemes ahead of the RNAPs and negatively supercoiled ones behind it, a feature known as "Twin Supercoiled Domain" (TSD). The present work aims at deciphering into some more detail the torsional dynamics of circular DNA molecules being transcribed by RNAP enzymes. To this end, we performed Brownian Dynamics simulations with a specially designed coarse-grained model. Depending on the superhelical density of the DNA molecule and the ratio of RNAP's twist injection rate and rotational relaxation speed, simulations reveal a rich panel of behaviors, which sometimes differ markedly from the crude TSD picture. In particular, for sufficiently slow rotational relaxation speed, positively supercoiled plectonemes never form ahead of a RNAP that transcribes a DNA molecule with physiological negative supercoiling. Rather, negatively supercoiled plectonemes form almost periodically at the upstream side of the RNAP and grow up to a certain length before detaching from the RNAP and destabilizing rapidly. The extent to which topological barriers hinder the dynamics of TSDs is also discussed.

**SIGNIFICANCE :** It has been known for four decades that a RNA Polymerase enzyme which transcribes torsionally relaxed DNA favors the formation of positively supercoiled plectonemes downstream of its position and negatively supercoiled ones upstream of it. How these plectonemes are born, how they evolve and eventually vanish, has however received little attention. We used coarse-grained modeling and Brownian Dynamics simulations to provide theoretical insight into this question, for both torsionally relaxed DNA and negatively supercoiled (bacterial) DNA. The contact maps obtained from these simulations reveal a rich dynamics, which depend essentially on the superhelical density of the DNA molecule and the torsional relaxation speed of the RNA Polymerase. Twin Supercoiled Domains are certainly NOT static features but instead very dynamic ones.

**INTRODUCTION**

Bending and torsional rigidity are responsible for geometrical deformations of double-stranded DNA molecules known as "twist" ($Tw$) and "writhe" ($Wr$) [1]. $Tw$ is equal to the number of times the strands of the double helix coil around each other and $Wr$ to the number of times the double helix coils around itself [2]. The sum of twist and writhe, $Lk = Tw + Wr$, which is called the "linking number", remains constant for circular DNA molecules, that is, for molecules which form closed loops, like the genomic DNA molecule of most bacteria. "Superhelical density", defined according to $\sigma = (Lk - Lk_0)/Lk_0$, where $Lk_0$ is the linking number of the torsionally relaxed DNA molecule, also remains constant for circular DNA molecules as long as their integrity is preserved. Torsional stress is, generally speaking, not distributed homogeneously along DNA molecules, which are instead organized into independent topological domains, whose linking numbers are not affected by changes in the torsional state of neighboring domains [3,4]. For example, the genome of *E. coli* is composed of several hundreds of topological domains with average size ≈10000 base pairs (bp) [5]. Neighboring domains are separated from each other by macromolecules which act as topological barriers, in the sense that they prevent equilibration of the linking numbers of the two domains. A well characterized example of topological barrier consists of a pair of LacI repressor proteins [6-9], which bridge double-stranded DNA, thereby forming a loop which is topologically isolated from the rest of the chromosome [6,10-12]. Distribution of torsional stress inside each topological domain is furthermore altered by several families of molecules, like RNA Polymerase (RNAP) enzymes [13-15], which catalyze the transcription of genes into messenger RNA. Transcription requires that double-stranded DNA be locally and transiently opened [16-18], which results in an increase of $Tw$ downstream of the RNAP enzyme and a decrease of $Tw$ upstream [19-22]. When the DNA molecule is torsionally relaxed ($\sigma = 0$) and the transcribing RNAP experiences sufficiently large rotational drag, for example when it is bound to a long nascent RNA molecule, the variations in $Tw$ lead to positively supercoiled plectonemes ahead of transcribing RNAPs and negatively supercoiled plectonemes behind them, a feature known as "Twin Supercoiled Domain" [19,23-25]. To be complete, let us mention that RNAPs are able to generate values of the superhelical density up to about $\sigma = -0.12$ [26,27], that is, four times the superhelical density of protein-bound DNA in living *E. coli* cells [28] and twice that of naked DNA *in vitro* [29]. *In vivo*, $\sigma$ is however maintained in a range compatible with physiological functions and cell survival through the recruitment of Type I topoisomerase enzymes upstream of transcribing RNAPs and Type II topoisomerase enzymes (gyrases) downstream of the RNAPs. These enzymes break temporarily one (respectively, two) DNA strands and relax negative (respectively, positive) supercoils [30-33]. By doing so, they modify the value of the linking number of the topological domain in which transcription is taking place, but they do not alter the linking number of neighboring domains located beyond topological barriers.

The present work deals with the torsional dynamics of circular DNA molecules being transcribed by RNAP enzymes. It is striking that relatively little is known about how plectonemes actually form and evolve, although the TSD model has been proposed almost four decades ago and is evoked as soon as transcribing RNAPs come into play. A coarse-grained bead-and-spring model, which was recently introduced to determine under which conditions DNA-bridging proteins may act as topological barriers [34,35], was modified to take transcribing RNAPs properly into account and its properties were investigated through Brownian Dynamics (BD) simulations. Simulations revealed a rich panel of behaviors, depending in particular on the superhelical density of the DNA molecule and the ratio of RNAP's twist injection rate and rotational relaxation speed. The influence of topological barriers on the dynamics of the plectonemes was also investigated, but not that of topoisomerases, whose modeling is left for future work. Since the present work describes in detail the time evolution of DNA contact maps during transcription, it is anticipated that it will prove helpful in deciphering experimental results obtained with Chromosomal Conformation Capture set-ups [36].

**METHODS**

*DNA molecule*

Circular DNA molecules are modeled as circular chains of $n = 2880$ beads with radius $a = 1.0$ nm separated at equilibrium by a distance $l_0 = 2.5$ nm. Each bead represents 7.5 bp, so that the chain of 2880 beads represents a 21600 bp DNA molecule. Associated to each bead $k$ are a vector $\mathbf{r}_k$, which describes the position of bead $k$ in the space-fixed frame, and a body-fixed orthogonal frame $(\mathbf{u}_k, \mathbf{f}_k, \mathbf{v}_k)$, which describes the torsional state of the molecule at bead $k$ [37]. $\mathbf{u}_k = (\mathbf{r}_{k+1} - \mathbf{r}_k) / \|\mathbf{r}_{k+1} - \mathbf{r}_k\|$ is the unit vector pointing from bead $k$ to bead $k+1$, whereas unit vectors $\mathbf{f}_k$ and $\mathbf{v}_k$ are orthogonal to

the chain at bead $k$. Knowledge of frames $(\mathbf{u}_k, \mathbf{f}_k, \mathbf{v}_k)$ and $(\mathbf{u}_{k+1}, \mathbf{f}_{k+1}, \mathbf{v}_{k+1})$ allows for the computation of the variation of the torsional angle between beads $k$ and $k+1$, $\Phi_{k+1} - \Phi_k$, as described in [37].

The potential energy of the DNA molecule, $E_{\text{DNA}}$, consists of four terms

$$E_{\text{DNA}} = \frac{h}{2}\sum_{k=1}^{n}(l_k - l_0)^2 + \frac{g}{2}\sum_{k=1}^{n}\theta_k^2 + \frac{\tau}{2}\sum_{k=1}^{n}(\Phi_{k+1} - \Phi_k)^2 + q^2\sum_{k=1}^{n-4}\sum_{m=k+4}^{n}H(\|\mathbf{r}_k - \mathbf{r}_m\| - 2a), \quad (1)$$

which describe the stretching, bending, torsional, and electrostatic energy of the DNA chain, respectively. By convention, $\mathbf{r}_{k+n} \equiv \mathbf{r}_k$ and $\Phi_{k+n} \equiv \Phi_k$. In Eq. (1), $l_k = \|\mathbf{r}_{k+1} - \mathbf{r}_k\|$ denotes the distance between successive beads $k$ and $k+1$ and $\theta_k$ the angle between vectors $\mathbf{r}_{k+1} - \mathbf{r}_k$ and $\mathbf{r}_k - \mathbf{r}_{k-1}$. The stretching rigidity was set to $h = 100\, k_B T / l_0^2$, where $T = 295$ K, because this value ensures that the variations of the distance between successive beads remain small enough when reasonably large integration time steps are used [38]. In contrast, the bending rigidity was deduced from the known persistence length of DNA, $\xi = 50$ nm, according to $g = \xi k_B T / l_0 = 20\, k_B T$. Torsion forces and momenta were computed as described in [37], except where otherwise stated. The value of the torsional rigidity, $\tau = 25\, k_B T$, ensures that the writhe contribution $\langle Wr \rangle$ accounts for approximately 2/3 of the linking number difference $\Delta Lk$ at equilibrium [39], see Figure S1 of [34]. Finally, the electrostatic energy is expressed as a sum of repulsive Debye-Hückel terms with hard core. Function $H(r)$ is defined according to

$$H(r) = \frac{1}{4\pi\varepsilon r}\exp\left(-\frac{r}{r_D}\right), \quad (2)$$

where $\varepsilon = 80\,\varepsilon_0$ is the dielectric constant of the buffer and $r_D = 1.07$ nm the Debye length inside the buffer. This value of the Debye length corresponds to a concentration of monovalent salt of 100 mM, which is the value that is generally assumed for the cytoplasm of bacterial cells. $q = -3.52\,\bar{e}$, where $\bar{e}$ is the absolute charge of the electron, is the value of the electric charge, which is placed at the center of each DNA bead. This value was deduced from Manning's counterion condensation theory [40,41]. Together with the small equilibrium distance between the centers of two DNA beads ( $l_0 = 2.5$ nm) and the large value of the stretching rigidity ($h = 100 \, k_\mathrm{B} T / l_0^2$), the repulsive Debye-Hückel terms with hard core ensure that two DNA strands cannot cross each other, which is of outmost importance in the present work.

*Overdamped Langevin equations*

The dynamics of the system was investigated by integrating numerically overdamped Langevin equations. Practically, the updated positions and torsion angles of DNA bead $k$ at time step $i+1$ were computed from the positions and torsion angles at time step $i$ according to

$$\mathbf{r}_k^{(i+1)} = \mathbf{r}_k^{(i)} + \frac{\Delta t}{6\pi\eta a}\mathbf{F}_k^{(i)} + \sqrt{\frac{2\,k_B T\,\Delta t}{6\pi\eta a}}\,\mathbf{x}_k^{(i)}, \qquad (3)$$

$$\Phi_k^{(i+1)} = \Phi_k^{(i)} + \frac{\tau\Delta t}{4\pi\eta a^2 l_0}(\Phi_{k+1}^{(i)} - 2\Phi_k^{(i)} + \Phi_{k-1}^{(i)}) + \sqrt{\frac{2\,k_B T\,\Delta t}{4\pi\eta a^2 l_0}}\,X_k^{(i)}, \qquad (4)$$

where the $\mathbf{F}_k^{(i)}$ are vectors of inter-particle forces arising from the total potential energy of the system, $T = 298$ K is the temperature of the system, $\mathbf{x}_k^{(i)}$ and $X_k^{(i)}$ are vectors of random numbers extracted from a Gaussian distribution of mean 0 and variance 1, $\eta = 0.00089$ Pa s is the viscosity of the buffer at 298 K, and $\Delta t = 5$ ps is the integration time step.

Updated vectors $\mathbf{u}_k^{(i+1)}$ of the body-fixed frames were then computed according to $\mathbf{u}_k^{(i+1)} = (\mathbf{r}_{k+1}^{(i+1)} - \mathbf{r}_k^{(i+1)})/\|\mathbf{r}_{k+1}^{(i+1)} - \mathbf{r}_k^{(i+1)}\|$, whereas vectors $\mathbf{f}_k^{(i+1)}$ and $\mathbf{v}_k^{(i+1)}$ were computed in a slightly different (and more accurate) way compared to [37], that is, by transforming $\mathbf{f}_k^{(i)}$ and $\mathbf{v}_k^{(i)}$ according to two successive rotations $(\mathbf{N}_k, \varphi_k)$ and $(\mathbf{u}_k^{(i+1)}, \Phi_k^{(i+1)} - \Phi_k^{(i)})$, where

$$\mathbf{N}_k = \frac{\mathbf{u}_k^{(i)} \times \mathbf{u}_k^{(i+1)}}{\|\mathbf{u}_k^{(i)} \times \mathbf{u}_k^{(i+1)}\|} \qquad (5)$$

$$\varphi_k = \sin^{-1}\left(\|\mathbf{u}_k^{(i)} \times \mathbf{u}_k^{(i+1)}\|\right). \qquad (6)$$

$\left(\mathbf{N}_k, \varphi_k\right)$ is the rotation that transforms $\mathbf{u}_k^{(i)}$ into $\mathbf{u}_k^{(i+1)}$. This scheme was used instead of Eq. (25) of [37], because it allows for a more natural modeling of transcribing RNAPs (see Section *RNA Polymerase*).

After each integration step, the excess of twist $\Delta T\text{w}$, the writhe $W\text{r}$, and the linking number difference $\Delta L\text{k} = \Delta T\text{w}+W\text{r}$ were computed as described in [37]. For the computation of $L\text{k}_0$ and $\sigma$, it was assumed that the two strands twist around the helical axis once every 10.5 base pairs in torsionally relaxed DNA. The time evolution of $\Delta L\text{k}$ was continuously monitored to ascertain that the crossing of two DNA strands, which would result in the abrupt variation of $\Delta L\text{k}$ by $\pm 2$, never occurs. This is illustrated in the top plot of Figure S1 of the Supporting Material.

*RNA Polymerase*

A transcribing RNAP enzyme bound to bead $\gamma$ of the DNA chain is taken into account by modifying the overdamped Langevin equations and the body-fixed frame update rules for this particular bead. More precisely, $\mathbf{r}_\gamma^{(i+1)}$ and $\Phi_\gamma^{(i+1)}$ are computed according to

$$\mathbf{r}_\gamma^{(i+1)} = \mathbf{r}_\gamma^{(i)} + \frac{\Delta t}{6\pi\eta A}\mathbf{F}_\gamma^{(i)} + \sqrt{\frac{2\,k_B T\,\Delta t}{6\pi\eta A}}\,\mathbf{x}_\gamma^{(i)}, \tag{7}$$

$$\Phi_\gamma^{(i+1)} = \Phi_\gamma^{(i)} + \frac{\tau\Delta t}{8\pi\eta A^3}(\Phi_{\gamma+1}^{(i)} - 2\Phi_\gamma^{(i)} + \Phi_{\gamma-1}^{(i)}) + \sqrt{\frac{2\,k_B T\,\Delta t}{8\pi\eta A^3}}\,X_\gamma^{(i)}, \tag{8}$$

instead of Eqs. (3) and (4), meaning that bead $\gamma$ behaves as a sphere of hydrodynamic radius $A \gg a$ instead of a cylinder of radius $a$ and length $l_0$. Unless otherwise stated, $A$ was set to 6 nm, which is the approximate dimension of a single RNAP enzyme [13,42]. In addition, vectors $\mathbf{f}_\gamma^{(i+1)}$ and $\mathbf{v}_\gamma^{(i+1)}$ are obtained by transforming $\mathbf{f}_\gamma^{(i)}$ and $\mathbf{v}_\gamma^{(i)}$ according to rotations $\left(\mathbf{N}_\gamma, \varphi_\gamma\right)$ and $\left(\mathbf{u}_\gamma^{(i+1)}, \Phi_\gamma^{(i+1)} - \Phi_\gamma^{(i)} - \Omega\Delta t\right)$, instead of $\left(\mathbf{N}_\gamma, \varphi_\gamma\right)$ and $\left(\mathbf{u}_\gamma^{(i+1)}, \Phi_\gamma^{(i+1)} - \Phi_\gamma^{(i)}\right)$. The additional angle $-\Omega\Delta t$ represents the energy-consuming action of the RNAP, which for positive rotational speed $\Omega$ actively winds the DNA molecule for beads $k > \gamma$ ("downstream") and unwinds it for beads $k < \gamma$ ("upstream").

$\Omega$ is taken as a free parameter, which is varied from 0.045 to 9.0 rad µs$^{-1}$ to investigate various dynamical regimes.

Since each bead represents 7.5 bp and the two strands of torsionally relaxed DNA twist around the helical axis once every 10.5 bp, translation of the RNAP is taken into account by incrementing $\gamma$ by $\mp 1$ each time the body-fixed frames in the neighborhood of the RNAP have rotated by $\pm(7.5/10.5)(2\pi)$. For this purpose, the time evolution of $\Phi_{\gamma+\Delta\gamma}$ was computed according to Eq. (4) for different values of $\Delta\gamma$. After checking that the results do not depend significantly on the choice of $\Delta\gamma$, this parameter was set to $\Delta\gamma = 10$. Many simulations were actually launched twice, that is both with and without RNAP translation, in order to discriminate between the effects of twist injection and RNAP translation.

*Topological barriers*

According to the model of topological barrier proposed in [35], the action of the barrier on bridged DNA beads $\alpha$ and $\beta$ is described by the potential energy function

$$E_{\text{TB}} = \frac{h}{2}(d_{\alpha\beta} - d_{\alpha\beta}^0)^2 + K(\mathbf{f}_\alpha \cdot \mathbf{f}_\beta + \mathbf{v}_\alpha \cdot \mathbf{v}_\beta), \quad (9)$$

where the first term describes the bond between beads $\alpha$ and $\beta$, whereas the second term ensures that the linking number remains constant in each loop. $d_{\alpha\beta} = \|\mathbf{r}_\beta - \mathbf{r}_\alpha\|$ denotes the distance between the centers of beads $\alpha$ and $\beta$, and the value $d_{\alpha\beta}^0 = 4$ nm is small enough to prevent DNA segments from crossing the line between the centers of beads $\alpha$ and $\beta$. $K$ was set to $1000 \, k_{\text{B}}T$, that is 40 times the torsional rigidity $\tau$. The contributions of the second term in the right-hand side of Eq. (9) to the torque $T_k$ and the force $\mathbf{F}_k$ felt by bead $k$ write [35]

$$T_\alpha :+ K(\mathbf{f}_\alpha \cdot \mathbf{v}_\beta - \mathbf{v}_\alpha \mathbf{f}_\beta), \quad (10)$$

$$T_\beta :- K(\mathbf{f}_\alpha \cdot \mathbf{v}_\beta - \mathbf{v}_\alpha \mathbf{f}_\beta), \quad (11)$$

$$\mathbf{F}_\alpha :- \frac{K}{l_\alpha}(\mathbf{u}_\alpha \cdot \mathbf{f}_\beta)\mathbf{f}_\alpha - \frac{K}{l_\alpha}(\mathbf{u}_\alpha \cdot \mathbf{v}_\beta)\mathbf{v}_\alpha, \quad (12)$$

$$\mathbf{F}_{\alpha+1} :+ \frac{K}{l_\alpha}(\mathbf{u}_\alpha \cdot \mathbf{f}_\beta)\mathbf{f}_\alpha + \frac{K}{l_\alpha}(\mathbf{u}_\alpha \cdot \mathbf{v}_\beta)\mathbf{v}_\alpha, \tag{13}$$

$$\mathbf{F}_\beta :- \frac{K}{l_\beta}(\mathbf{f}_\alpha \cdot \mathbf{u}_\beta)\mathbf{f}_\beta - \frac{K}{l_\beta}(\mathbf{v}_\alpha \cdot \mathbf{u}_\beta)\mathbf{v}_\beta, \tag{14}$$

$$\mathbf{F}_{\beta+1} :+ \frac{K}{l_\beta}(\mathbf{f}_\alpha \cdot \mathbf{u}_\beta)\mathbf{f}_\beta + \frac{K}{l_\beta}(\mathbf{v}_\alpha \cdot \mathbf{u}_\beta)\mathbf{v}_\beta, \tag{15}$$

The torques in Eqs. (10) and (11) replace the torques exerted by beads $\alpha \pm 1$ on bead $\alpha$ and those exerted by beads $\beta \pm 1$ on bead $\beta$. This model is referred to as "Model I" in the remainder of this paper.

In the present work, we also used a second model, according to which the action of the barrier on beads $\alpha$ and $\beta$ is described by the potential energy function

$$E_{\text{TB}} = \frac{h}{2}(d_{\alpha\beta} - d_{\alpha\beta}^0)^2 + K\,\mathbf{w}_{\alpha\beta} \cdot (\mathbf{f}_\beta - \mathbf{f}_\alpha), \tag{16}$$

where $\mathbf{w}_{\alpha\beta} = (\mathbf{r}_\beta - \mathbf{r}_\alpha)/\|\mathbf{r}_\beta - \mathbf{r}_\alpha\|$. The contributions of the second term in the right-hand side of Eq. (16) to the torque $T_k$ and the force $\mathbf{F}_k$ felt by bead $k$ write

$$T_\alpha :+ K\,\mathbf{w}_{\alpha\beta} \cdot \mathbf{v}_\alpha, \tag{17}$$

$$T_\beta :- K\,\mathbf{w}_{\alpha\beta} \cdot \mathbf{v}_\beta, \tag{18}$$

$$\mathbf{F}_\alpha :+ \frac{K}{l_\alpha}(\mathbf{w}_{\alpha\beta} \cdot \mathbf{u}_\alpha)\mathbf{f}_\alpha + \frac{K}{d_{\alpha\beta}}(\mathbf{f}_\beta - \mathbf{f}_\alpha) - \frac{K}{d_{\alpha\beta}}(\mathbf{w}_{\alpha\beta} \cdot (\mathbf{f}_\beta - \mathbf{f}_\alpha))\mathbf{w}_{\alpha\beta}, \tag{19}$$

$$\mathbf{F}_{\alpha+1} :- \frac{K}{l_\alpha}(\mathbf{w}_{\alpha\beta} \cdot \mathbf{u}_\alpha)\mathbf{f}_\alpha, \tag{20}$$

$$\mathbf{F}_\beta :- \frac{K}{l_\beta}(\mathbf{w}_{\alpha\beta} \cdot \mathbf{u}_\beta)\mathbf{f}_\beta - \frac{K}{d_{\alpha\beta}}(\mathbf{f}_\beta - \mathbf{f}_\alpha) + \frac{K}{d_{\alpha\beta}}(\mathbf{w}_{\alpha\beta} \cdot (\mathbf{f}_\beta - \mathbf{f}_\alpha))\mathbf{w}_{\alpha\beta}, \tag{21}$$

$$\mathbf{F}_{\beta+1} :+ \frac{K}{l_\beta}(\mathbf{w}_{\alpha\beta} \cdot \mathbf{u}_\beta)\mathbf{f}_\beta. \tag{22}$$

As for Model I, the torques in Eqs. (17) and (18) replace the torques exerted by beads $\alpha \pm 1$ on bead $\alpha$ and those exerted by beads $\beta \pm 1$ on bead $\beta$. The model of Eq. (16) is referred to as "Model II" in the remainder of this paper.

The linking number difference in each loop, $\Delta Lk^{(1)}$ and $\Delta Lk^{(2)}$, computed as described in [35], were constantly monitored in order to ascertain that topological barriers adequately play their role. An example is shown in the middle and bottom plots of Figure S1 of the Supporting Material.

The differences between Models I and II will be discussed in detail in Section *TSDs in the presence of topological barriers*.

**RESULTS**

*Rotational speed of the body-fixed frames*

The system was first investigated in the absence of topological barriers and RNAP translation, that is for a circular DNA chain with a single RNAP at position $\gamma = n/2 = 1440$. The superhelical density of the DNA chain was either $\sigma = 0$ (torsionally relaxed DNA) or $\sigma = -0.063$, which corresponds approximately to the superhelical density of naked bacterial DNA *in vitro* [29]. *A* was set either to 6 nm, which is the approximate dimension of a single RNAP enzyme [13,42], or to 6 μm, in order to model a much less mobile RNAP bound to a long nascent RNA molecule [24]. Simulations were launched for 8 different values of Ω ranging from 0.045 to 9.0 rad μs$^{-1}$.

The first conclusion that can be drawn from the simulations is that, after a transient period which lasts up to several tens of ms, the system reaches a steady state where internal vectors $(\mathbf{f}_k, \mathbf{v}_k)$ rotate on average at uniform and constant rotational speed around $\mathbf{u}_k$. This can be deduced from the plots of $<\Phi_{k+1} - \Phi_k>$ as a function of $k$, where $<\Phi_{k+1} - \Phi_k>$ represents the time average of the variation of the torsional angle between beads $k$ and $k+1$. For example, Figure S2 of the Supporting Material shows the variations of $<\Phi_{k+1} - \Phi_k>$ for $A = 6$ nm, $\sigma = 0$ (top plot) or $\sigma = -0.063$ (bottom plot), and 4 different values of Ω. The gray horizontal dotted lines represent the value of $<\Phi_{k+1} - \Phi_k>$ in the absence of RNAP and the green vertical dot-dashed lines the position of the RNAP ($\gamma = n/2 = 1440$). Remember also that beads $k = 1$ and $k = n$ are connected. It is clear from Figure S2 that $<\Phi_{k+1} - \Phi_k>$ decreases linearly with $k$, except for the discontinuity at $\gamma$. According to Eq. (4), this linear decrease indicates that internal vectors $(\mathbf{f}_k, \mathbf{v}_k)$ rotate on average at uniform speed

$$\omega = \frac{\tau}{4\pi\eta a^2 l_0} < \overline{\Phi_{k+1} - 2\Phi_k + \Phi_{k-1}} > , \qquad (23)$$

around $\mathbf{u}_k$, where $<\overline{\Phi_{k+1} - 2\Phi_k + \Phi_{k-1}}>$ denotes the average of $\Phi_{k+1} - 2\Phi_k + \Phi_{k-1}$ over both time and $k$ (except for the discontinuity at $k = \gamma$). Evolution of $-\omega$ as a function of $\Omega$ is shown in Figure 1 for simulations performed with $\sigma = 0$ and $A = 6$ μm ($\otimes$ symbols), $\sigma = 0$ and $A = 6$ nm (+ symbols), and $\sigma = -0.063$ and $A = 6$ nm ($\times$ symbols). Dot-dashed lines are guidelines for the eyes, which highlight the fact that $-\omega$ increases proportionally to $\Omega$ over all the investigated range. Moreover, values obtained with $\sigma = -0.063$ are very close to those obtained with $\sigma = 0$, meaning that $\omega$ is essentially independent of $\sigma$. Finally, Figure 1 indicates that larger values of $|\omega|$ may be obtained by increasing either $\Omega$ or $A$. Varying $\Omega$ or $A$, and consequently $|\omega|$, represents a straightforward way to address experimental situations where the rotational relaxation speed of the transcription machinery varies with respect to the rate of injection of twist in the DNA molecule. In the present work, we arbitrarily chose to vary $\Omega$ rather than $A$ in order to investigate different regimes, and consequently performed all further simulations with $A = 6$ nm. Still, we note that $-\omega$ is about 30 times smaller than $\Omega$ for $A = 6$ nm and 7.5 times smaller for $A = 6$ μm, which indicates that most part of the additional rotation $-\Omega \Delta t$ imposed to bead $\gamma$ at each time step is immediately balanced by the restoring torsional forces and torques exerted by neighboring DNA beads. $\Omega$ is consequently a model parameter without obvious counterpart in the real world. The important parameter with that respect is the rotation speed of the internal bases, $\omega$.

An additional point worth noting is that the activity of the RNAP modifies the steady state contributions of the mean excess of twist $\langle \Delta T\text{w}\rangle$ and the mean writhe $\langle Wr\rangle$ to the linking number difference $\Delta L\text{k}$ of supercoiled DNA, as was already noted in [35] using a slightly different model. Indeed, it is seen in Figure S3 of the Supporting Material that for $\sigma = -0.063$ the writhe contribution $\langle Wr\rangle / \Delta L\text{k}$ increases significantly for $\Omega \geq 0.90$ rad μs$^{-1}$ and becomes as large as ≈0.80 for $\Omega \geq 4.5$ rad μs$^{-1}$, against 0.70 in the absence of RNAP.

*TSDs at $\sigma = 0$ in the absence of topological barriers*

Simulations performed with $\sigma = 0$ and increasing values of Ω indicate that clear TSDs appear only for $\Omega \geq 4.5$ rad μs$^{-1}$. Indeed, for $\Omega = 2.25$ rad μs$^{-1}$ only few, unstable and short plectonemes are observed close to the active RNAP, while plectonemes are basically absent from simulations performed with $\Omega \leq 0.9$ rad μs$^{-1}$.

The dynamics of the TSDs which form for $\Omega \geq 4.5$ rad μs$^{-1}$ was investigated by plotting maps showing the probability of contact between two beads of the DNA chain. These maps were computed by checking every $2.10^5$ time steps (1 μs time intervals) whether the centers of beads *k* and *m* are separated by less than 10 nm and computing the corresponding contact probability over 100 successive time intervals (100 μs time windows). Density maps showing the logarithm of the contact probability as a function of *k* and *m* (hereafter called *contact maps*) were subsequently plotted for each time window. A representative contact map obtained in the course of a simulation with $\sigma = 0$, $\gamma = n/2 = 1440$ and $\Omega = 9.0$ rad μs$^{-1}$ in the absence of topological barriers and RNAP translation is shown in Figure 2 (top). In this figure, the vertical and horizontal green dashed lines indicate the position of the RNAP. Plectonemes appear as straight segments oriented perpendicular to the main diagonal of the figure, like the segment labeled (p). Since the RNAP increases supercoiling downstream of its position ($k > \gamma$) and decreases supercoiling upstream ($k < \gamma$), segments parallel to (p) and located in the top right quadrant denote positively supercoiled plectonemes, whereas those located in the lower left quadrant denote negatively supercoiled plectonemes. In the contact map shown in Figure 2 (top), the RNAP therefore sits at the foot of a long negatively supercoiled plectoneme (beads $645 \leq k \leq 1440$, line of contacts (p)) and, simultaneously, at the foot of a shorter positively supercoiled plectoneme (beads $1440 \leq k \leq 1550$, line of contacts (p*)). This is illustrated in Figure 3, where the RNAP is shown as a red sphere, the plectoneme associated with contacts (p) as a green tube and the plectoneme associated with contacts (p*) as a blue tube. Besides plectonemes, another line of contacts (labeled (a)) is clearly seen in Figure 2 (top). This line of contacts is due to the fact that the long negatively supercoiled plectoneme (p) located upstream of the RNAP winds around a tract of the DNA chain (beads $1550 \leq k \leq 1675$, orange tube in Figure 3) located downstream of the RNAP, just after the short positively supercoiled plectoneme (p*). Reciprocally, the short positively supercoiled plectoneme (p*) located downstream of the RNAP winds around a DNA tract (beads $605 \leq k \leq 645$, yellow tube in Figure 3) located upstream of the RNAP, just before the long negatively supercoiled plectoneme (p).

The sequence of contact maps obtained in the course of a simulation with $\sigma = 0$, $\gamma = n/2 = 1440$ and $\Omega = 9.0$ rad $\mu s^{-1}$ in the absence of topological barriers and RNAP translation is assembled in Movie S1 of the Supporting Material. The movie consists of 2000 frames displayed at 20 fps and encompassing a real time window of 200 ms. It is seen in Movie S1 that plectonemes grow alternatively and quasi-periodically upstream and downstream of the RNAP, with the RNAP sitting at one foot of the plectoneme and the plectoneme winding around a tract of the DNA chain located on the other side of the RNAP. After some time, the long plectoneme detaches from the RNAP, unwinds from the DNA tract and moves further away from the RNAP, while another plectoneme starts growing on the other side of the RNAP. Quite interestingly, the plectoneme which moves away from the RNAP becomes progressively unstable and dissolves before reaching the tract of the DNA chain opposite to the RNAP. The reason is that $<\Phi_{k+1} - \Phi_k>$ vanishes in the neighborhood of $k = 1$ and $k = n$, that is opposite to $k = \gamma = n/2$ (see Figure S2 (top)). Since the torque exerted by bead $k+1$ on bead $k$ is equal to $\tau(\Phi_{k+1} - \Phi_k)$ [37], the average torque vanishes around $k = 1$ and $k = n$, which means that the conditions for long-lived plectonemes are not satisfied in this tract of the DNA chain. A complete sequence is illustrated in Figure 4 (top), where 5 contact maps separated by 4.0 ms are shown in different colors ranging from red to blue and superposed.

Plectoneme dynamics was further investigated by plotting the time evolution of the local writhe

$$Wr(k, l_w) = \frac{1}{4\pi} \int_{k-l_w}^{k} \int_{k}^{k+l_w} \frac{(\mathbf{r}_1 - \mathbf{r}_2) \cdot (d\mathbf{r}_1 \times d\mathbf{r}_2)}{|\mathbf{r}_1 - \mathbf{r}_2|^3} \ . \qquad (24)$$

$Wr(k, l_w)$ is the signed counterpart of the "local unsigned writhing" discussed in [43-45] and provides a measure of the self-crossings of segment $[k - l_w, k + l_w]$. As suggested in [45], $l_w$ was set to $3\xi$, that is 150 nm or 60 beads. Of interest to us is the fact that the apex of a plectoneme corresponds to an extremum of $Wr(k,l_w)$ [43,44]. The growth and dynamics of plectonemes is therefore reflected in the evolution of blue lines (apex of negatively supercoiled plectonemes) and red lines (apex of positively supercoiled plectonemes) in Figure S4 (bottom) of the Supporting Material, which shows the time evolution of $Wr(k,l_w)$ for a simulation performed with $\sigma = 0$ and $\Omega = 9.0$ rad μs$^{-1}$. In this figure, the yellow line indicates the position of the RNAP. Black vertical segments indicate those moments when a long plectoneme detaches from one side of the RNAP while another plectoneme starts growing on the other side. The length of the black segments represents the half-length of the detaching plectonemes. Figure S4 (bottom) highlights the fact that for $\Omega = 9.0$ rad μs$^{-1}$ plectonemes detach almost periodically from the RNAP when they reach a rather uniform half-length of about 310 beads, that is slightly more than 2000 bp. The interval of ±310 beads is shown as a set of two dashed horizontal lines surrounding the position of the RNAP.

RNAP translation was finally taken into account in the simulations, in order to check to what extent it interferes with the dynamics arising from twist injection. The sequence of contact maps obtained in the course of a simulation with $\sigma = 0$ and $\Omega = 9.0$ rad μs$^{-1}$, with RNAP translation but in the absence of topological barriers, is assembled in Movie S2 of the Supporting Material. Movie S2 consists of 1000 frames displayed at 20 fps and encompassing a real time window of 100 ms. It clearly shows that the translation speed of the RNAP is significantly larger than the speed at which plectonemes grow and move away from the RNAP, which hinders the growth of long positively supercoiled plectonemes at the downstream side of the translating RNAP. Inspection of Movie S2 and Figure 5 (top), which shows the time evolution of $Wr(k,l_w)$ for all DNA beads *k*, confirms that RNAP translation modifies significantly the dynamics of TSDs. Indeed, positively supercoiled plectonemes do not form at the downstream side of the translating RNAP, but rather ahead of it, sometimes as far as 700 or 800 beads (about 5000 bp) away. As the RNAP catches up the foot of one these plectonemes, a line of contacts similar to line (a) of Figure 2 (top) usually forms, as if the positively supercoiled plectoneme were growing. However, the RNAP translates more rapidly than the plectoneme grows, so that the RNAP "digests" progressively the plectoneme, up to the moment when the plectoneme almost disappears. Subsequently, a negatively supercoiled plectoneme grows at the upstream side of the RNAP, which is usually accompanied again by the formation of a (a)-type line of contacts. After a while, the negatively supercoiled plectoneme detaches from the RNAP, separates more widely from it and dissolves progressively, as in the absence of RNAP translation. Noteworthy, the positively supercoiled plectonemes which form ahead of the RNAP and the negatively supercoiled plectonemes which detach from it are significantly shorter (half-length of about 150 beads, that is about 1000 bp) than those observed in the absence of RNAP translation (half-length of about 310 beads, that is about 2000 bp). This is compensated by the fact that the number of plectonemes that are formed per unit time is about twice larger in the presence of RNAP translation. The two key steps displaying the (a)-type line of contacts are clearly seen in red and blue in Figure 4 (bottom), where 3 contact maps separated by 2.0 ms are shown in different colors and superposed.

*TSDs at $\sigma = -0.063$ in the absence of topological barriers*

The simple picture that a transcribing RNAP creates positively supercoiled plectonemes downstream of its position and negatively supercoiled ones upstream does not hold when the DNA molecule is itself supercoiled in the absence of RNAP. For example, if $\sigma < 0$, negatively supercoiled plectonemes may be present at any position along the DNA molecule in the absence of RNAP, so that a transcribing RNAP is expected to strengthen the negative supercoiling of plectonemes located upstream of its position and destabilize instead those located downstream. In the present section, we discuss in detail the effects of a transcribing RNAP on a circular DNA molecule with $\sigma = -0.063$.

Simulations performed with $\sigma = -0.063$ and increasing values of $\Omega$ indicate that clear TSDs appear only for $\Omega \geq 4.5$ rad $\mu s^{-1}$, that is exactly as for simulations performed with $\sigma = 0$. This means that the threshold for the formation of TSDs is independent of $\Omega$ in the investigated range. Below this threshold, the main effect of a RNAP transcribing a closed DNA chain with $\sigma = -0.063$ is to promote the rearrangement of the existing short plectonemes into a single long plectoneme, with the RNAP sitting at (or close to) the apex of the plectoneme. This is clearly seen in Movie S3 of the Supporting Material (1600 frames displayed at 20 fps, 160 ms real time window), which shows the sequence of contact maps obtained in the course of a simulation with $\sigma = -0.063$ and $\Omega = 0.9$ rad $\mu s^{-1}$. The mere difference, which is observed upon taking RNAP translation into account, is that the length of the plectoneme at the apex of which the RNAP sits decreases for increasing values of $\Omega$. For example, for $\Omega = 0.9$ rad $\mu s^{-1}$ the half-length of the plectoneme reduces to about 450 beads (about 3000 bp). Active molecular motors sitting at the apex of a plectoneme have already been reported, both experimentally [25,46] and according to simulations performed with models which differ from the present one [35,47].

On the other hand, the dynamics of TSDs observed for $\sigma = -0.063$ and $\Omega \geq 4.5$ rad $\mu s^{-1}$ differs significantly from that of TSDs at $\sigma = 0$, as can be checked in Movie S4 of the Supporting Material (2000 frames displayed at 20 fps, 200 ms real time), which shows the sequence of contact maps obtained in the course of a simulation with $\sigma = -0.063$, $\gamma = n/2 = 1440$ and $\Omega = 9.0$ rad $\mu s^{-1}$, in the absence of topological barriers and RNAP translation. As for $\sigma = 0$, long plectonemes grow almost periodically at the RNAP but, it contrast with $\sigma = 0$, the plectonemes are all negatively supercoiled and grow only upstream of the RNAP, instead of growing alternately on both sides of it. The reason for this difference is of course that the action of the transcribing RNAP opposes the initial supercoiling of plectonemes located downstream but strengthens that of plectonemes located upstream. As for $\sigma = 0$, the long plectoneme detaches from the RNAP after reaching a certain length, unwinds from the DNA tract, moves further away from the RNAP and destabilizes rapidly. A complete sequence is illustrated in Figure 6 (top), where 5 contact maps separated by 5.0 ms are shown in different colors ranging from red to blue and superposed.

A representative contact map and the corresponding DNA conformation are shown in Figure 2 (bottom) and Figure 7, respectively. As for $\sigma = 0$, the RNAP sits simultaneously at one foot of the long plectoneme (beads $520 \leq k \leq 1440$, line of contacts (p) in Figure 2 (bottom), green tube in Figure 7) and one foot of a shorter plectoneme located on its other side (beads $1440 \leq k \leq 1540$, line of contacts (p*) in Figure 2 (bottom), blue tube in Figure 7). The long plectoneme winds around a tract of the DNA chain located just after the short plectoneme (beads $1540 \leq k \leq 1682$, line of contacts (a) in Figure 2 (bottom), orange tube in Figure 7), whereas the short plectoneme winds around a tract of the DNA chain located just before the long plectoneme (beads $480 \leq k \leq 520$, yellow tube in Figure 7). Compared to $\sigma = 0$, an additional line of contacts, labeled (b) in Figure 2 (bottom), is observed in most frames of Movie S4. This line of contacts is due to a DNA tract located downstream of the RNAP (beads $1682 \leq k \leq 2550$, cyan tube in Figure 7) winding around a second, shorter DNA tract located upstream of the RNAP (beads $1 \leq k \leq 480$, black tube in Figure 7).

Figure S5 (bottom) of the Supporting Material, which shows the time evolution of the local writhe $Wr(k, l_w)$ for a simulation performed with $\sigma = -0.063$ and $\Omega = 9.0$ rad μs$^{-1}$ in the absence of RNAP translation, indicates that plectonemes detach from the RNAP at an average half-length of about 570 beads ( about 4000 bp). It also highlights the fact that the latency between the moments when a plectoneme reaches the maximum length and when it detaches from the RNAP may be quite long. This is for example the case for the third event in Figure S5 (bottom), whereas no significant latency was observed for $\sigma = 0$ (Figure S4 (bottom)).

Last but not least, simulations suggest that RNAP translation interferes to a significantly lesser extent with the dynamics arising from twist injection for $\sigma = -0.063$ than for $\sigma = 0$. This can be checked in Movie S5 of the Supporting Material (800 frames shown at 20 fps, real time window of 800 ms), as well as Figure 6 (bottom), which were obtained from a simulation with $\sigma = -0.063$ and $\Omega = 9.0$ rad μs$^{-1}$, with RNAP translation but in the absence of topological barriers. The movie clearly shows that the forward motion of the RNAP does not hinder the formation of negatively supercoiled plectonemes at its upstream side. Still, comparison of Figure S5 (bottom) and Figure 5 (bottom), which shows the time evolution of $Wr(k, l_w)$ for all DNA beads $k$, indicates that RNAP translation has three significant effects on the dynamics of the plectonemes. First, RNAP translation enhances the rate of formation of plectonemes by about one order of magnitude, probably because it allows for a faster relaxation of the plectoneme being formed. However, the increased formation rate is also due in part to the fact that the latency between the moments when a plectoneme reaches the maximal length and when it detaches from the RNAP is suppressed. Moreover, the increased formation rate comes at the expense of a somewhat reduced half-length of the plectonemes, namely about 370 bp (slightly less than 3000 bp), against 570 bp (about 4000 bp) in the absence of RNAP translation. Finally, the line of contacts labeled (b) in Figure 2 (bottom) is no longer observed for a translating RNAP (Movie S5), so that DNA conformations look like Figure 3 instead of Figure 7.

We note that it is not clear whether "Twin Supercoiled Domain" is a correct designation for the plectonemes whose dynamics is discussed in the present subsection, because of the almost total lack of positively supercoiled plectonemes ahead of the RNAP. This expression, as well as the TSD acronym, will nevertheless be used in a loose manner in the remainder of this paper to designate the plectonemes which are born as a consequence of a transcribing RNAP.

*TSDs in the presence of topological barriers*

Let us now consider a RNAP which transcribes a DNA chain that is divided into two loops by a Model I topological barrier. The second term in the right-hand side of Eq. (9) ensures that vectors $\mathbf{f}_\alpha$ and $\mathbf{v}_\alpha$ of the internal basis at bead $\alpha$ remain antiparallel to vectors $\mathbf{f}_\beta$ and $\mathbf{v}_\beta$ of the internal basis at bead $\beta$, respectively. As a consequence, the internal bases at beads $\alpha$ and $\beta$ are forced to rotate synchronously, which ensures that the excess of twist remains constant in each loop [34,35]. Moreover, the tract of the DNA chain surrounding bead $\alpha$ cannot rotate with respect to the tract of DNA chain surrounding bead $\beta$, which implies that transfers between intra-loop writhe and inter-loop entanglement are blocked [34,35]. The conjunction of these two restrictions ensures that the linking number remains constant in each loop [34,35].

Simulations were performed with a topological barrier bridging beads $\alpha = n/2 = 1440$ and $\beta = n = 2880$, thus dividing the DNA chain into two equal loops, and a RNAP located at $\gamma = 3n/4 = 2160$, that is at the center of one loop. Two different values of DNA superhelical density ($\sigma = 0$ and $\sigma = -0.063$) and four different values of $\Omega$ (0.9, 2.25, 4.5 and 9.0 rad μs$^{-1}$) were systematically investigated. Figure S6 of the Supporting Material shows the variations at steady state of $<\Phi_{k+1} - \Phi_k>$ as a function of $k$ for $\sigma = 0$ (top plot) and $\sigma = -0.063$ (bottom plot). The green vertical dot-dashed lines represent the position of the RNAP ($\gamma = 3n/4 = 2160$) and the red ones the positions of the two beads connected by the topological barrier ($\alpha = n/2 = 1440$ and $\beta = n = 2880$). Except for the discontinuities at $\alpha$, $\beta$ and $\gamma$, $<\Phi_{k+1} - \Phi_k>$ clearly decreases linearly with $k$, as was also the case in Figure S2 for simulations performed without topological barrier. As discussed in Section *Rotational speed of the body-fixed frames*, this indicates that vectors $(\mathbf{f}_k, \mathbf{v}_k)$ of the internal bases rotate on average at uniform and constant rotational speed $\omega$ around vectors $\mathbf{u}_k$ (Eq. (23)). The values of $\omega$ computed from the simulations with a Model I topological barrier are equal to 75% of those computed from simulations without topological barrier (Figure 1), whatever the value of $\Omega$. Most importantly, simulations reveal that the presence of a Model I topological barrier does not alter significantly the dynamics of TSDs described above, the reason being that plectonemes formed in the loop containing the transcribing RNAP are able to cross the topological barrier and penetrate the other loop, and *vice versa*.

We also considered a RNAP which transcribes a DNA chain that is divided into two loops by a Model II topological barrier. The second term in the right-hand side of Eq. (16) is minimum when $\mathbf{f}_\beta$ is antiparallel to both $\mathbf{f}_\alpha$ and $\mathbf{r}_\beta - \mathbf{r}_\alpha$. As a consequence, $(\mathbf{f}_\alpha, \mathbf{v}_\alpha)$ and $(\mathbf{f}_\beta, \mathbf{v}_\beta)$ can no longer rotate around $\mathbf{u}_\alpha$ and $\mathbf{u}_\beta$, respectively. This term additionally forbids rotation of the tract of DNA chain surrounding bead $\alpha$ with respect to the tract of DNA chain surrounding bead $\beta$, which ensures that the linking number remains constant in each loop [34,35].

As for Model I, simulations were performed with a Model II topological barrier bridging beads $\alpha = n/2 = 1440$ and $\beta = n = 2880$ and a RNAP located at $\gamma = 3n/4 = 2160$. Two different values of DNA superhelical density ($\sigma = 0$ and $\sigma = -0.063$) and four different values of $\Omega$ (0.9, 2.25, 4.5 and 9.0 rad μs$^{-1}$) were systematically investigated. Figure S7 of the Supporting Material shows the evolution at steady state of $<\Phi_{k+1} - \Phi_k>$ as a function of $k$ for $\sigma = 0$ (top plot) and $\sigma = -0.063$ (bottom plot). In contrast with Figure S2 and Figure S6, $<\Phi_{k+1} - \Phi_k>$ remains constant in each DNA segment, meaning that vectors $(\mathbf{f}_k, \mathbf{v}_k)$ no longer rotate around vectors $\mathbf{u}_k$. Indeed, since rotation of $(\mathbf{f}_\alpha, \mathbf{v}_\alpha)$ and $(\mathbf{f}_\beta, \mathbf{v}_\beta)$ is blocked by the topological barrier, the system is expected to stall as soon as the torque exerted by the DNA beads located on both sides of the transcribing RNAP balances the effect of the active angle $-\Omega \Delta t$. The major consequence is of course that the dynamics of TSDs in the presence of a Model II topological barrier can only consist of a short transient followed by fluctuations around the stalled conformation, whatever the value of $\sigma$. This conclusion is fully confirmed by the simulations, as can be checked in Movie S6 of the Supporting Material (1000 frames shown at 20 fps, real time window of 100 ms), which shows the sequence of contact maps obtained with $\sigma = 0$ and $\Omega = 9.0$ rad μs$^{-1}$. For $\sigma = 0$, the stalled conformation consists of a positively supercoiled plectoneme located downstream of the RNAP and a negatively supercoiled one located upstream. In contrast, for $\sigma = -0.063$, the stalled conformation consists of a single negatively supercoiled plectoneme located upstream of the RNAP.

*Evolution of plectonemes' properties with $\Omega$*

A detailed study of the evolution of plectonemes' properties with $\Omega$ would require many more simulations than launched in the present work. Still, some general trends can be gained from the comparison of simulations performed with $\Omega = 4.5$ rad $\mu s^{-1}$ and $\Omega = 9.0$ rad $\mu s^{-1}$. For example, comparison of the top and bottom plots in Figure S4 and Figure S5 shows that the length of fully grown plectonemes does not depend significantly on $\Omega$ in the investigated range in the absence of RNAP translation. In addition, the average slope of red and blue lines in the top plot of Figure S4 and Figure S5 is about one half of the average slope in the bottom plot, which means that the speed at which plectonemes grow and move away from the RNAP increases linearly with $\Omega$. Finally, the number of full-length plectonemes that are formed per unit time appears to increase rapidly (supra-linearly) with $\Omega$, the reason being that a larger number of aborted growth events, during which small plectonemes detach too early from the RNAP, take place for $\Omega = 4.5$ rad $\mu s^{-1}$ compared to $\Omega = 9.0$ rad $\mu s^{-1}$. Comparison of movies obtained with $\Omega = 4.5$ rad $\mu s^{-1}$ and $\Omega = 9.0$ rad $\mu s^{-1}$ indicates that these conclusions remain likely true when RNAP translation is taken into account, probably because the speed of RNAP translation and the rate of twist injection are linearly related.

**DISCUSSION**

In this section, the results obtained in the present work are discussed in connection with previous related work.

The conformation of the DNA chain shown in Figure 3 is reminiscent of the so-called hyper-plectonemes, which were investigated in detail in [48]. Indeed, the AFM images of DNA molecules longer than 14 kbp reported in [48], which form higher-order filaments composed of several intertwined individual plectonemes, look like the green DNA plectoneme winding around the orange DNA tract in Figure 3. There exist, however, two fundamental differences between the hyper-plectonemes of [48] and the conformation displayed in Figure 3. Indeed, hyper-plectonemes consist of plectonemes winding around plectonemes, whereas the conformation displayed in Figure 3 consists of a plectoneme winding around double-stranded DNA. Moreover, the formation of hyper-plectonemes requires effective DNA-DNA attraction, which may for instance be achieved by incubating supercoiled DNA in the presence of increasing concentrations of DNA-bridging nucleoid proteins, like H-NS, FIS of HU [48]. The conformation in Figure 3 was instead obtained with repulsive DNA-DNA interactions, but it necessitates active injection of torsional stress by the RNAP. Still, the resemblance between the two conformations is appealing and suggests that both mechanisms may contribute similarly to the compaction of the bacterial nucleoid [48].

Concerning an altogether different matter, the interactions between RNAPs and loop-extruding SMC complexes were investigated in [49]. It was shown that the speed of condensin translocation is slowed down when the condensin interacts with a RNAP complex, with the slowing down being significantly more important when the condensin and the RNAP collide head-to-head rather than head-to-tail [49]. It may be worth pointing out that the present work provides a temptative explanation for this result. Indeed, our simulations suggest that the two sides of a RNAP transcribing supercoiled DNA are not "equivalent". More precisely, it was shown in Section *Results*, that the long plectoneme which forms on the upstream side of the RNAP winds around a DNA tract located downstream of the RNAP (green and orange tubes in Figure 3), whereas the short plectoneme which forms on the downstream side of the RNAP winds around a DNA tract located upstream of the RNAP (blue and yellow tubes in Figure 3). Stated in other words, the upstream side of the RNAP is significantly less cluttered than its downstream side. A condensin complex which catches up a RNAP moving in the same direction approaches it from its upstream side, that is the cleaner one. Once the small plectoneme and the RNAP have overtaken the condensin ring, the long plectoneme is pulled by the RNAP without much further resistance. In contrast, a condensin complex which collides head-to-head with a translating RNAP encounters first the long plectoneme. One can imagine that the forward motion of the condensin lets the long plectoneme unwind from the DNA and form a blob, which may in turn seriously impede the crossing of the condensin and the RNAP. This is a point which eventually deserves further investigation.

Let us conclude by comparing the results discussed in the present paper and those reported in the only other paper related to the dynamics of TSDs we are aware of [50]. The work in [50] is based on a finer-grained model with single nucleotide resolution, which fully accounts for the double helical structure of DNA and allows for explicit denaturation of the two strands. Moreover, RNAP was explicit taken into account in the form of a rigid body composed of a ring and a crossbar which passes between the two DNA strands and opens them. Simulations were performed with 1000 bp or 2000 bp DNA chains, which were initially torsionally relaxed ($\sigma = 0$). A topological barrier was usually placed on the DNA tract opposite to the RNAP. The results in [50] must therefore be compared with the early stage of the simulations illustrated in Movie S2, Figure 4 (bottom) and Figure 5 (top) or in Movie S6. In spite of their huge differences, the two models agree strikingly well in predicting the main feature of the dynamics of plectonemes on torsionally relaxed DNA chains, that is the asymmetric formation of negatively and positively supercoiled plectonemes (simulations shown in Movie S2 and Movie S6 correspond to $v \approx 1.5 \times 10^{-3}$ bp$/\tau_{\text{Br}}$ in Figure S7 of [50]). Indeed, the former ones form and grow at the upstream side of the RNAP, whereas the latter ones form and grow quite far ahead of the RNAP. The present work indicates that this asymmetry is even more pronounced for negatively supercoiled DNA molecules, which are most common in prokaryotes, in the sense that only negatively supercoiled plectonemes form in the wake of the RNAP (see Movie S5, Figure 5 (bottom) and Figure 6 (bottom)).

**CONCLUSION**

A coarse-grained model and Brownian Dynamics simulations were used to decipher the torsional dynamics of a circular DNA molecule being transcribed by a RNAP enzyme. Simulations performed without topological barrier and for sufficiently large twist injection rates suggest that the crude TSD picture, according to which positively supercoiled plectonemes are generated ahead of the RNAP and negatively supercoiled ones behind it, is approximately satisfied for torsionally relaxed DNA molecules with, however, a significant asymmetry in birth places. Indeed, positively supercoiled plectonemes form up to several thousands bp ahead of the RNAP, whereas negatively supercoiled ones form directly at the upstream side of the RNAP. This asymmetry is even more pronounced (and the crude TSD picture is invalidated) for negatively supercoiled DNA molecules, which are most common in prokaryotes, in the sense that only negatively supercoiled plectonemes form at the upstream side of the RNAP. In both cases, negatively supercoiled plectonemes detach from the RNAP after reaching a size of several thousands bp and destabilize rapidly as their distance to the RNAP increases. Moreover, our simulations reveal that the transcribing RNAP is responsible for systematic arrays of DNA contacts, which extend several thousands bp away from it. Since Chromosome Conformation Capture techniques are nowadays able to image contacts at a resolution better than 1000 bp [51], and observation of the time evolution of contact maps might be feasible shortly [52], it is anticipated that the present work will prove helpful to decipher the internal organization of bacterial nucleoids. Finally, our simulations confirm that topological barriers perturb significantly the dynamics of Twin Supercoiled Domains. The degree of hindrance depends sensitively on the details of the topological barrier, but it is very likely that pairs of LacI repressor proteins [6-12], which probably work like Model II topological barriers, are able to stall the whole system after a short while. *In vivo*, this struggle is solved by topoisomerases, which relax supercoils both upstream and downstream of the RNAP. It may therefore be interesting to extend the coarse-grained model proposed here to take topoisomerases suitably into account.

**AUTHOR CONTRIBUTIONS**

This work is by a single author, who designed and performed research, analyzed results and wrote the paper.

**DELCARATION OF INTERESTS**

The author declares no competing interests.

**SUPPORTING MATERIAL**

Figures S1-S7. Movies S1-S6.

# References

1. Bates, A.D., and A. Maxwell. 2005.DNA topology. Oxford University Press, New York.

2. Cammack, R., and T. Atwood. 2006. Oxford dictionary of biochemistry and molecular biology. Oxford University Press, New York.

3. Worcel, A., and E. Burgi. 1972. On the structure of the folded chromosome of Escherichia coli. *J. Mol. Biol.* 71:127-47.

4. Sinden, R.R., and D.E. Pettijohn. 1981. Chromosomes in living Escherichia coli cells are segregated into domains of supercoiling. *Proc. Natl. Acad. Sci. USA*.78:224-228.

5. Postow, L., C.D. Hardy, J. Arsuaga, and N.R. Cozzarelli. 2004. Topological domain structure of the Escherichia coli chromosome. *Genes Dev.* 18:1766-1779.

6. Leng, F., B. Chen, and D.D. Dunlap. 2011. Dividing a supercoiled DNA molecule into two independent topological domains. *Proc. Natl. Acad. Sci. USA.* 108:19973-19978.

7. Müller-Hill, B. 1998. Some repressors of bacterial transcription. *Curr. Opin. Microbiol.* 1:145-151.

8. Fulcrand, G., S. Dages, X. Zhi, P. Chapagain, B.S. Gerstman, D. Dunlap, and F. Leng. 2016. DNA supercoiling, a critical signal regulating the basal expression of the lac operon in Escherichia coli. *Sci. Rep.* 6:19243.

9. Fulcrand, G., P. Chapagain, D. Dunlap, and F. Leng. 2016. Direct observation of a 91 bp LacI-mediated, negatively supercoiled DNA loop by atomic force microscope. *FEBS Lett.* 590:613-618.

10. Saldanha, R., P. Flanagan, and M. Fennewald. 1987. Recombination by resolvase is inhibited by lac repressor simultaneously binding operators between res sites. *J. Mol. Biol.* 196:505-516.

11. Wu, H.Y., and L.F. Liu. 1991. DNA looping alters local DNA conformation during transcription. *J. Mol. Biol.* 219:615-622.

12. Yan, Y., Y. Ding, F. Leng, D. Dunlap, and L. Finzi. 2018. Protein-mediated loops in supercoiled DNA create large topological domains. *Nucleic Acids Res.* 46:4417-4424.

13. Bae, B., A. Feklistov, A. Lass-Napiorkowska, R. Landick, and S.A. Darst. 2015. Structure of a bacterial RNA polymerase holoenzyme open promoter complex. *eLife.* 4:e08504.

14. Schafer, D.A., J. Gellest, M.P. Sheetz, and R. Landick. 1991. Transcription by single molecules of RNA polymerase observed by light microscopy. *Nature.* 352:444-448.

15. Yin, H., R. Landick, and J. Gelles. 1994. Tethered Particle Motion Method for Studying Transcript Elongation by a Single RNA Polymerase Molecule. *Biophys. J.* 67:2468-2478.

16. Wang, M.D., M.J. Schnitzer, H. Yin, R. Landick, J. Gelles, and S.M. Block. 1998. Force and velocity measured for single molecules of RNA polymerase. *Science.* 282:902-907.

17. Thomen, P., P.J. Lopez, U. Bockelmann, J. Guillerez, M. Dreyfus, and F. Heslot. 2008. T7 RNA polymerase studied by force measurements varying cofactor concentration. *Biophys. J.* 95:2423–2433.

18. Wuite, G.J.L., S.B. Smith, M. Young, D. Keller, and C. Bustamante. 2000. Single molecule studies of the effect of template tension on T7 DNA polymerase activity. *Nature.* 404:103-106.

19. Liu, L., and J.C. Wang. 1987. Supercoiling of the DNA template during transcription. *Proc. Natl. Acad. Sci. USA.* 84:7024-7027.

20. Deng, S., R.A. Stein, and N.P. Higgins. 2004. Transcription-induced barriers to supercoil diffusion in the Salmonella typhimurium chromosome. *Proc. Natl. Acad. Sci. USA.* 101:3398-3403.

21. Deng, S., R.A. Stein, and N.P. Higgins. 2005. Organization of supercoil domains and their reorganization by transcription. *Mol. Microbiol.* 57:1511-1521.

22. Kim, S., B. Beltran, I. Irnov, and C. Jacobs-Wagner. 2019. Long-distance cooperative and antagonistic RNA Polymerase dynamics via DNA supercoiling. *Cell.* 179:106–119.

23. Wu, H.Y., S. Shyy, J.C. Wang, and L.F. Liu. 1988. Transcription generates positively and negatively supercoiled domains in the template. *Cell*. 53:433-440.

24. Tsao, Y.P., H.Y. Wu, and L.F. Liu. 1989. Transcription-driven supercoiling of DNA: Direct biochemical evidence from in vitro studies. *Cell*. 56:111-118.

25. Janissen, R., R. Barth , M. Polinder, J. van der Torre, and C. Dekker. 2024. Single-molecule visualization of twin-supercoiled domains generated during transcription. *Nucleic Acids Res.* 52:1677–1687.

26. Samul, R., and F. Leng. 2007. Transcription-coupled hypernegative supercoiling of plasmid DNA by T7 RNA polymerase in Escherichia coli topoisomerase I-deficient strains. *J. Mol. Biol.* 374:925–935.

27. Leng, F., and R. McMacken. 2002. Potent stimulation of transcription-coupled DNA supercoiling by sequence-specific DNA-binding proteins. *Proc. Natl. Acad. Sci. USA*. 99:9139–9144.

28. Bliska, J.B., and N.R. Cozzarelli. 1987. Use of site-specific recombination as a probe of DNA structure and metabolism in vivo. *J. Mol. Biol.* 194:205-218.

29. Bauer, W.R. 1978. Structure and reactions of closed duplex DNA. *Annu. Rev. Biophys. Bioeng.* 7:287-313.

30. Reece, R.J., and A. Maxwell. 1991. DNA gyrase: structure and function. *Crit. Rev. Biochem. Mol. Biol.* 6:335–375.

31. McKie, S.J., K.C. Neuman, and A Maxwell. 2021. DNA topoisomerases: Advances in understanding of cellular roles and multi-protein complexes via structure-function analysis. *BioEssays*. 43:2000286.

32. Ahmed, W., C. Sala, S.R. Hegde, R.Kumar Jha, S.T. Cole, and V. Nagaraja. 2017. Transcription facilitated genome-wide recruitment of topoisomerase I and DNA gyrase. *PLoS Genet.* 13:e1006754.

33. Ma, J., and M.D. Wang. 2016. DNA supercoiling during transcription. *Biophys. Rev.* 8:S75-S87.

34. Joyeux, M., and I. Junier. 2020. Requirements for DNA-bridging proteins to act as topological barriers of the bacterial genome. *Biophys. J.* 119:1215-1225.

35. Joyeux, M. 2022. Models of topological barriers and molecular motors of bacterial DNA. *Mol. Simul.* 48:1688-1696.

36. McCord, R.P., N. Kaplan, and L. Giorgetti. 2020. Chromosome Conformation Capture and beyond: Toward an integrative view of chromosome structure and function. *Mol. Cell.* 77:688-708.

37. Chirico, G., and J. Langowski. 1994. Kinetics of DNA supercoiling studied by Brownian dynamics simulation. *Biopolymers*. 34:415-433.

38. Jian, H., A. Vologodskii, and T. Schlick. 1997. A combined wormlike-chain and bead model for dynamic simulations of long linear DNA. *J. Comp. Phys.* 136:168-179.

39. Boles, T.C., J.H. White, and N.R. Cozzarelli. 1990. Structure of plectonemically supercoiled DNA. *J. Mol. Biol.* 213:931-951.

40. Manning, G.S. 1969. Limiting laws and counterion condensation in polyelectrolyte solutions. I. Colligative properties. *J. Chem. Phys.* 51:924-933.

41. Oosawa, F. 1971. Polyelectrolytes. Marcel Dekker, New York.

42. Gnatt, A.L., P. Cramer, J. Fu, D.A. Bushnell, and R.D. Kornberg. 2001. Structural basis of transcription: An RNA Polymerase II elongation complex at 3.3 A resolution. *Science*. 292:1876-1882.

43. Vologodskii, A.V., S.D. Levene, K.V. Klenin, M. Frank-Kamenetskii, and N.R. Cozzarelli. 1992. Conformational and thermodynamic properties of supercoiled DNA. *J. Mol. Biol.* 227:1224-1243.

44. Klenin, K., and J. Langowski. 2000. Computation of writhe in modeling of supercoiled DNA. *Biopolymers*. 54:307–317.

45. Michieletto, D. 2016. On the tree-like structure of rings in dense solutions. *Soft Matter*. 12:9485-9500.

46. ten Heggeler-Bordier, B., W. Wahli, M. Adrian, A. Stasiak, and J. Dubochet. 1992. The apical localization of transcribing RNA-polymerases on supercoiled DNA prevents their rotation around the template. *EMBO J.* 11:667-672.

47. Racko, D., F. Benedetti, J. Dorier, and A. Stasiak. 2018. Transcription-induced supercoiling as the driving force of chromatin loop extrusion during formation of TADs in interphase chromosomes. *Nucleic Acids Res.* 46:1648-1660.

48. Japaridze, A., G. Muskhelishvili, F. Benedetti, A.F.M. Gavriilidou, R. Zenobi, P. De Los Rios, G. Longo, and G. Dietler. 2017. Hyperplectonemes: A higher order compact and dynamic DNA self-organization. *Nano Lett.* 17:1938−1948.

49. Brandão, H.B., P. Paul, A.A. van den Berg, D.Z. Rudner, X. Wang, and L.A. Mirny. 2019. RNA polymerases as moving barriers to condensing loop extrusion. *Proc. Natl. Acad. Sci. USA.* 116: 20489–20499.

50. Fosado, Y.A.G., D. Michieletto, C.A. Brackley, and D. Marenduzzo. 2021. Nonequilibrium dynamics and action at a distance in transcriptionally driven DNA supercoiling. *Proc. Natl. Acad. Sci. USA.* 118:e1905215118.

51. Cameron, C.J.F., J. Dostie, and M. Blanchette. 2020. HIFI: estimating DNA-DNA interaction frequency from Hi-C data at restriction-fragment resolution. *Genome Biol.* 21:11.

52. Dekker, J., A.S. Belmont, M. Guttman, V.O. Leshyk, J.T. Lis, S. Lomvardas, L.A. Mirny, C.C. O'Shea, P.J. Park, B. Ren, J.C. Ritland Politz, J. Shendure, and S. Zhong. 2017. The 4D nucleome project. *Nature*. 549:219-226.

**FIGURE CAPTIONS**

**Figure 1.** Evolution of $-\omega$, the rotational speed of internal bases, as a function of $\Omega$, the RNAP active rotational speed, in the absence of topological barriers and RNAP translation. Simulations were performed with $\sigma=0$ or $\sigma=-0.063$, $A=6$ nm or $A=6$ µm, and $\gamma=n/2=1440$.

**Figure 2.** Representative contact maps extracted from simulations with $\sigma=0$ (top) or $\sigma=-0.063$ (bottom), $\gamma=n/2=1440$ and $\Omega=9.0$ rad µs$^{-1}$ in the absence of topological barriers and RNAP translation. $k$ and $m$ are the indexes of DNA beads. The horizontal and vertical green dashed lines indicate the position of the RNAP.

**Figure 3.** Zoom-in on the DNA conformation corresponding to the contact map shown in Figure 2 (top) and highlighting the tracts of the DNA chain involved in the lines of contacts labeled (p), (p*) and (a). The plectoneme responsible for the line of contacts (p) is shown as a green tube and the one responsible for the line of contacts (p*) as a blue tube. The line of contacts (a) is due to the long green plectoneme winding around the tract of DNA shown as an orange tube.

**Figure 4.** Composite maps extracted from simulations with $\sigma=0$, $\gamma=n/2=1440$ and $\Omega=9.0$ rad µs$^{-1}$ in the absence of topological barriers. RNAP translation is neglected in the top plot. It is instead taken into account in the bottom plot. The horizontal and vertical dashed lines indicate the position of the RNAP. The top plot consists of the superposition of 5 contact maps separated by 4.0 ms and arranged in the following order: red, yellow, light green, dark green, and blue. The bottom plot consists of the superposition of 3 contact maps separated by 2.0 ms and arranged in the following order: red, light green, and blue.

**Figure 5.** Density plots showing the time evolution of the local writhe $Wr(k,l_w)$ for all DNA beads $k$, computed from simulations with $\Omega=9.0$ rad µs$^{-1}$ and $\sigma=0$ (top) or $\sigma=-0.063$ (bottom), with RNAP translation but in the absence of topological barriers. The yellow lines indicate the position of the translating RNAP.

**Figure 6.** Composite maps extracted from simulations with $\sigma = -0.063$, $\gamma = n/2 = 1440$ and $\Omega = 9.0$ rad μs$^{-1}$ in the absence of topological barriers. RNAP translation is neglected in the top plot. It is instead taken into account in the bottom plot. The horizontal and vertical dashed lines indicate the position of the RNAP. The top plot consists of the superposition of 5 contact maps separated by 5.0 ms and arranged in the following order: red, yellow, light green, dark green, and blue. The bottom plot consists of the superposition of 3 contact maps separated by 2.5 ms and arranged in the following order: red, light green, and blue.

**Figure 7.** Zoom-in on the DNA conformation corresponding to the contact map shown in Figure 2 (bottom) and highlighting the tracts of the DNA chain involved in the lines of contacts labeled (p), (p*), (a) and (b). The plectoneme responsible for the line of contacts (p) is shown as a green tube and the one responsible for the line of contacts (p*) as a blue tube. The line of contacts (a) is due to the long green plectoneme winding around the tract of DNA shown as an orange tube. The line of contacts (b) is due to the tract of DNA shown as a cyan tube winding around the tract of DNA shown as a black tube.

FIGURE 1

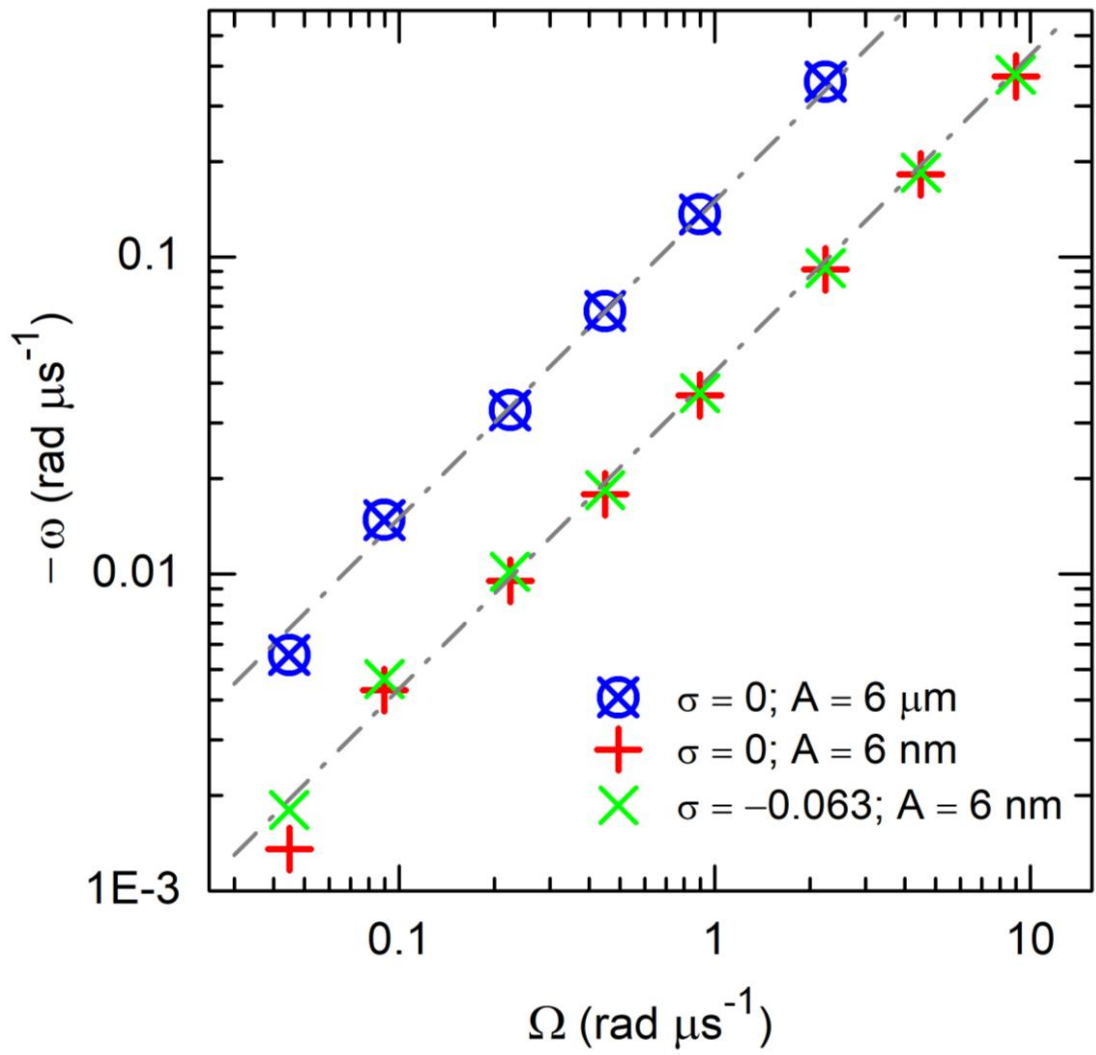

FIGURE 2

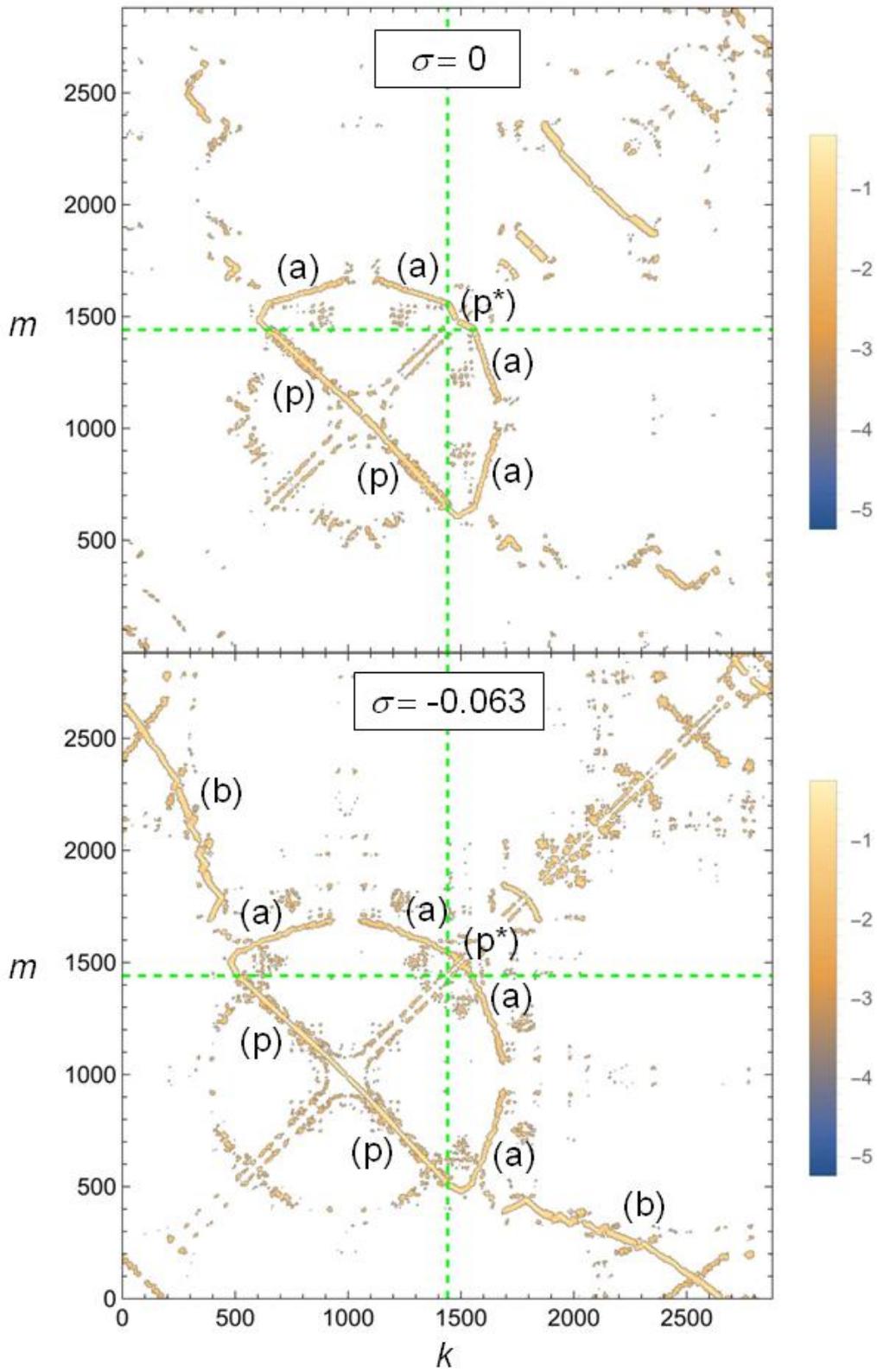

FIGURE 3

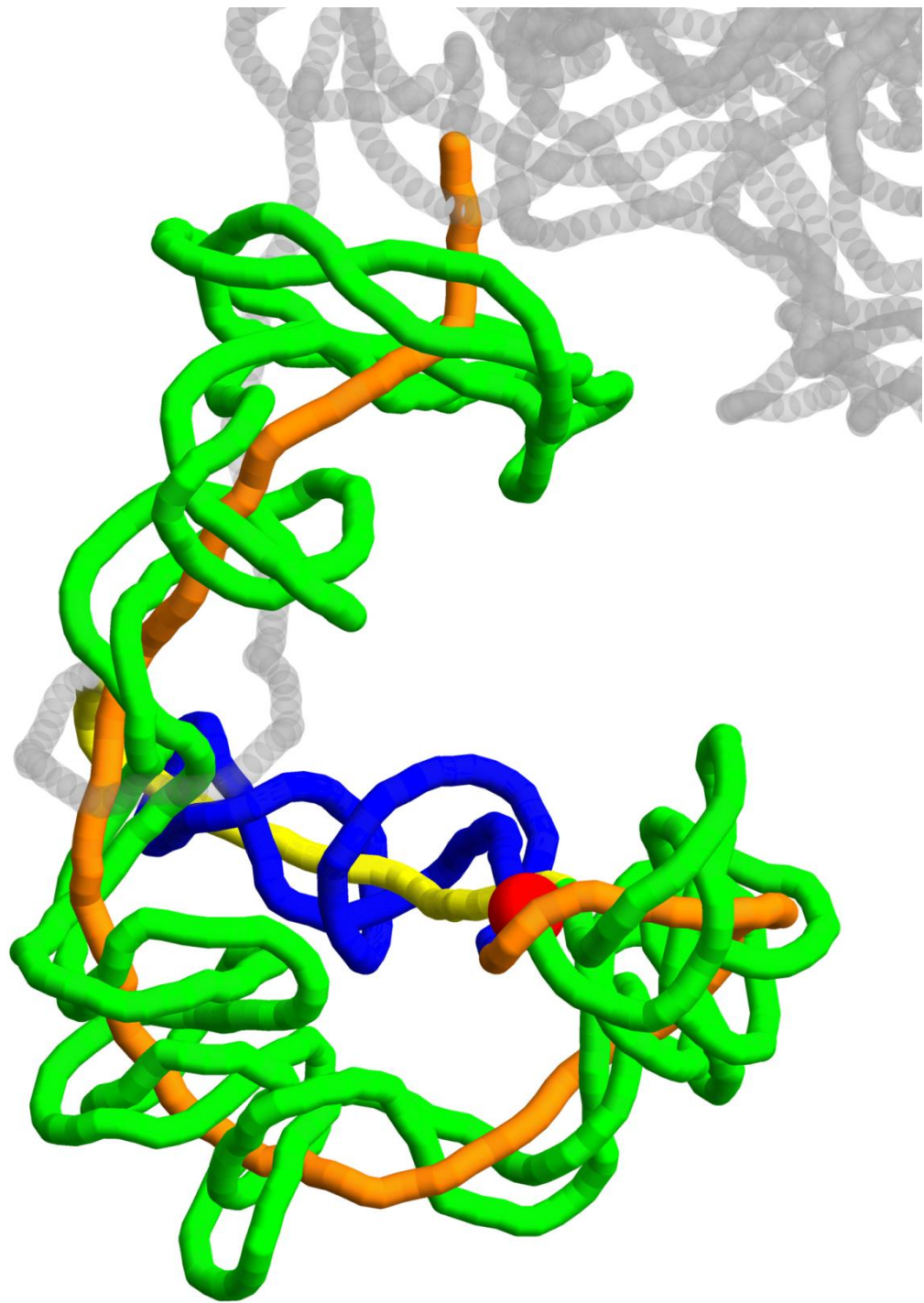

FIGURE 4

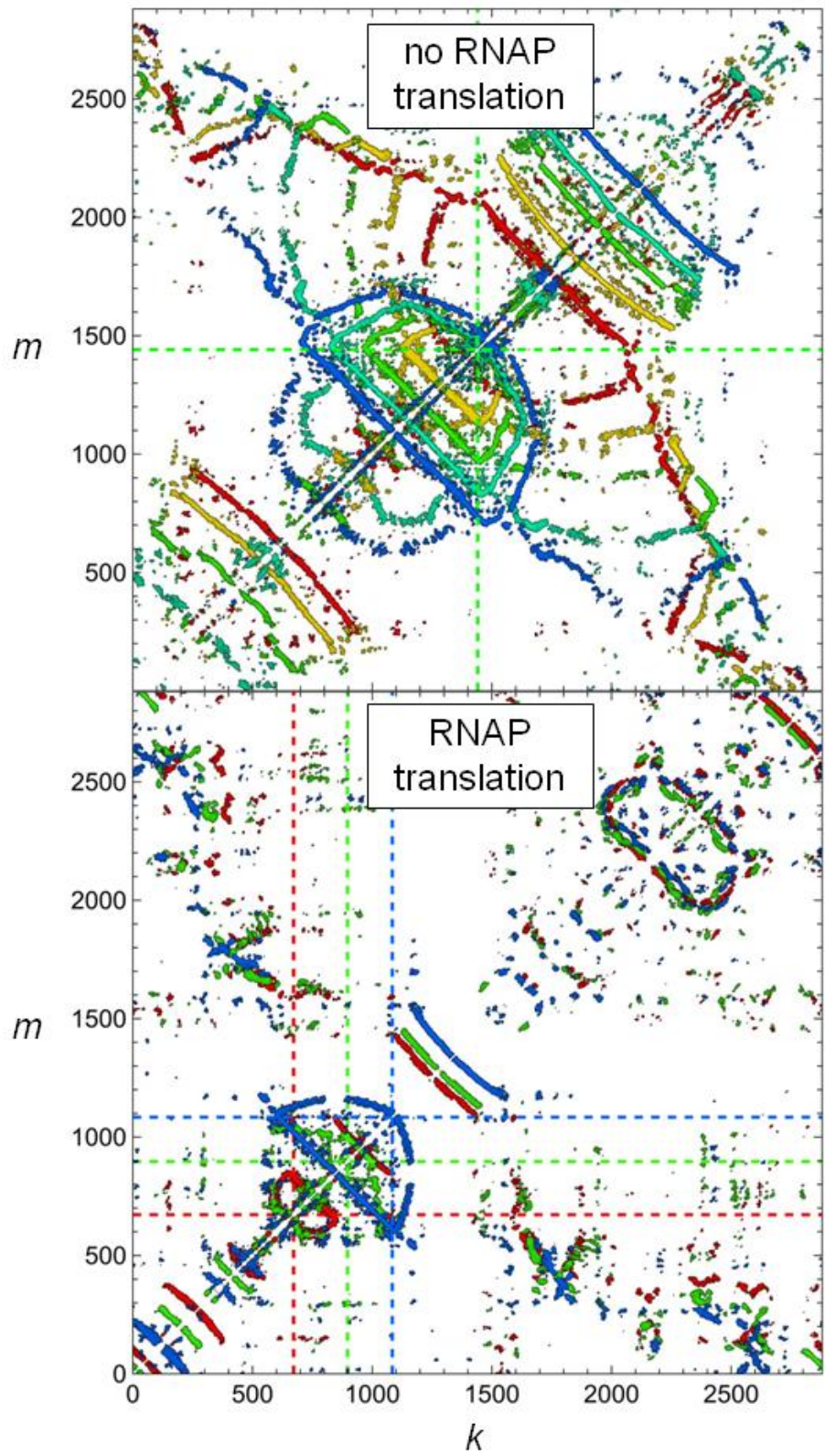

FIGURE 5

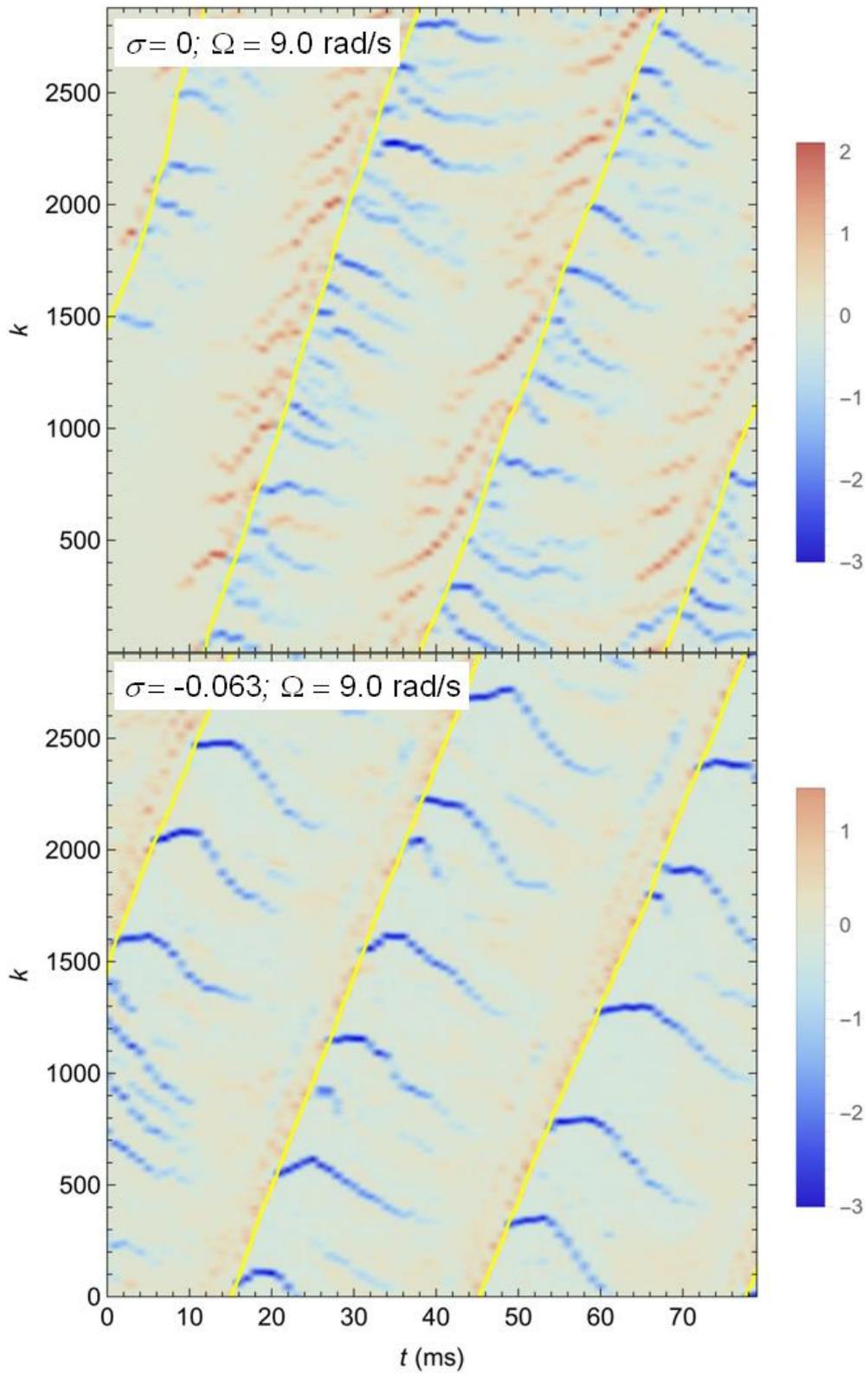

FIGURE 6

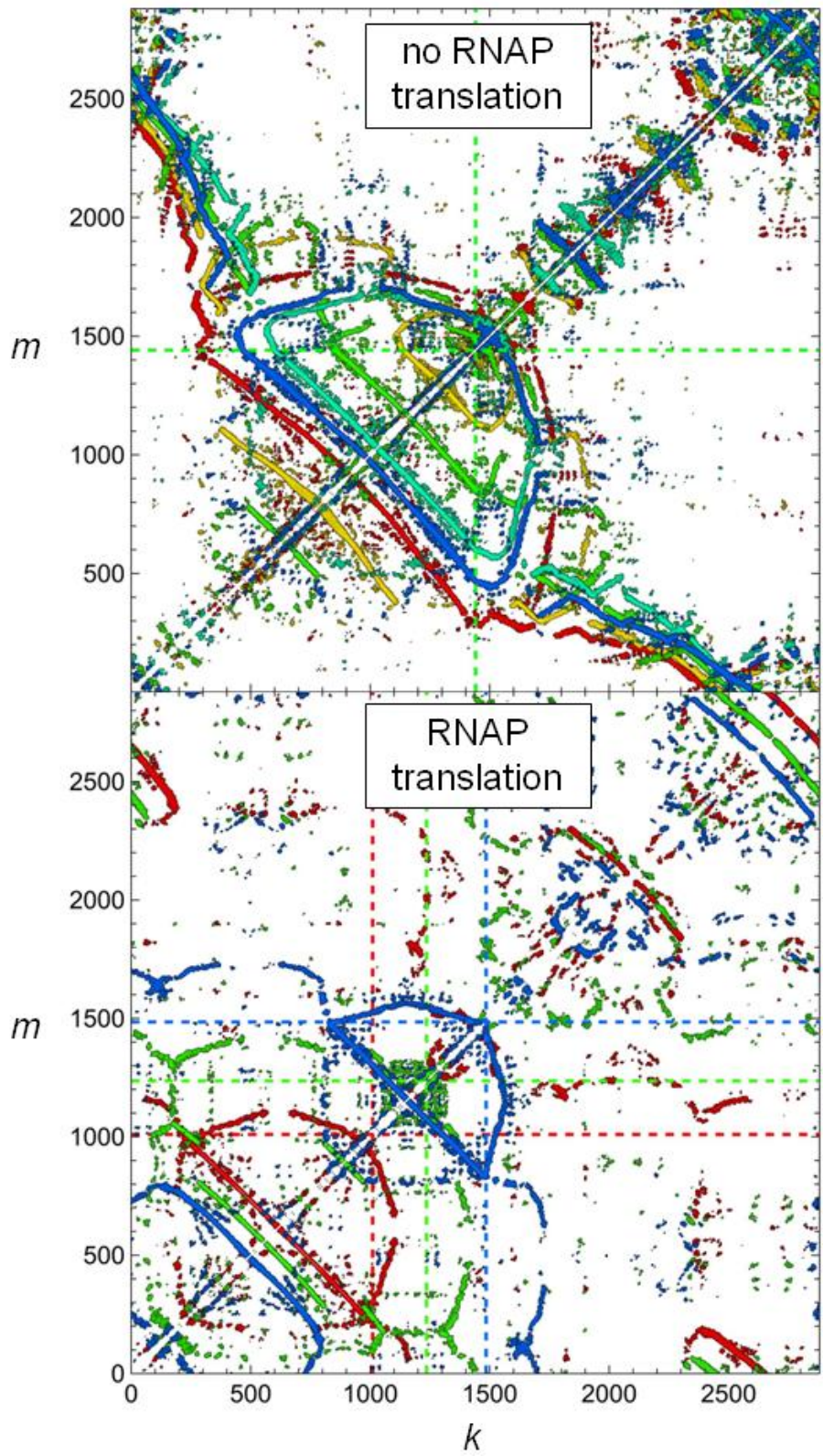

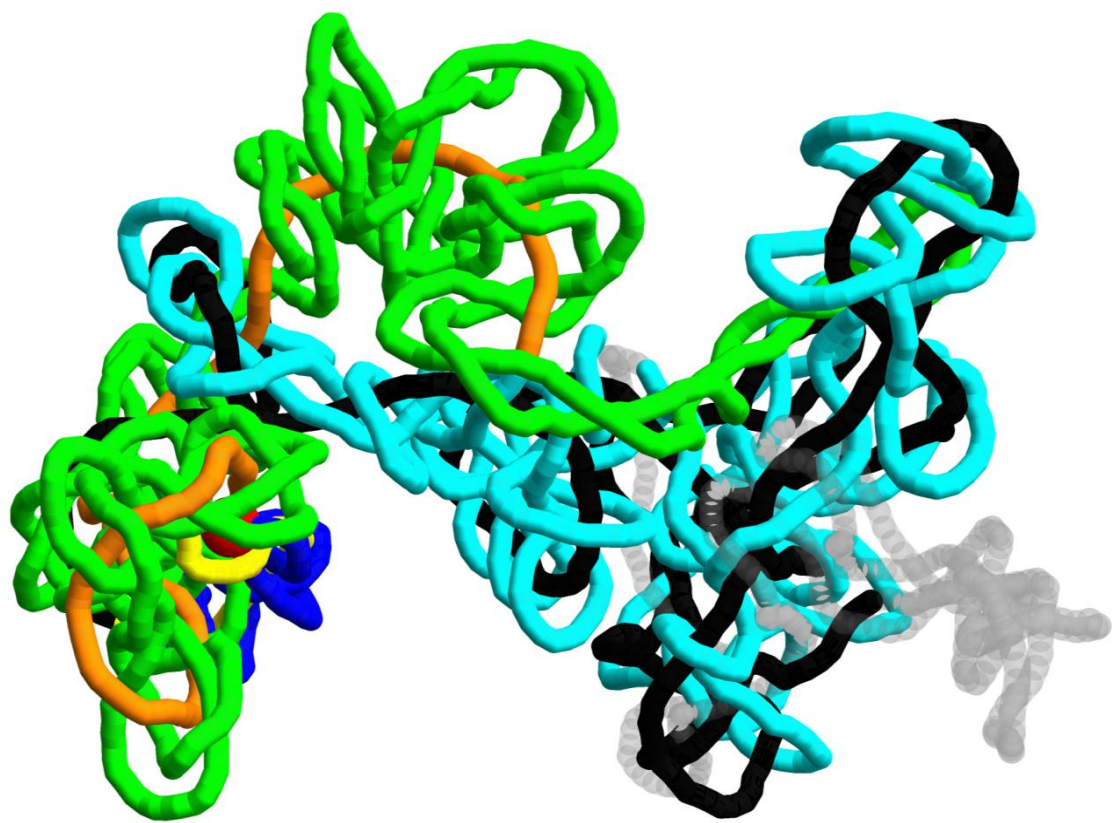

**Transcribing RNA polymerases :**
**Dynamics of Twin Supercoiled Domains**
- Supporting Material –

M. Joyeux
*Laboratoire Interdisciplinaire de Physique,*
*CNRS and Université Grenoble Alpes,*
*Grenoble, France*

- Figures S1 to S7

- Captions of Movies S1 to S6

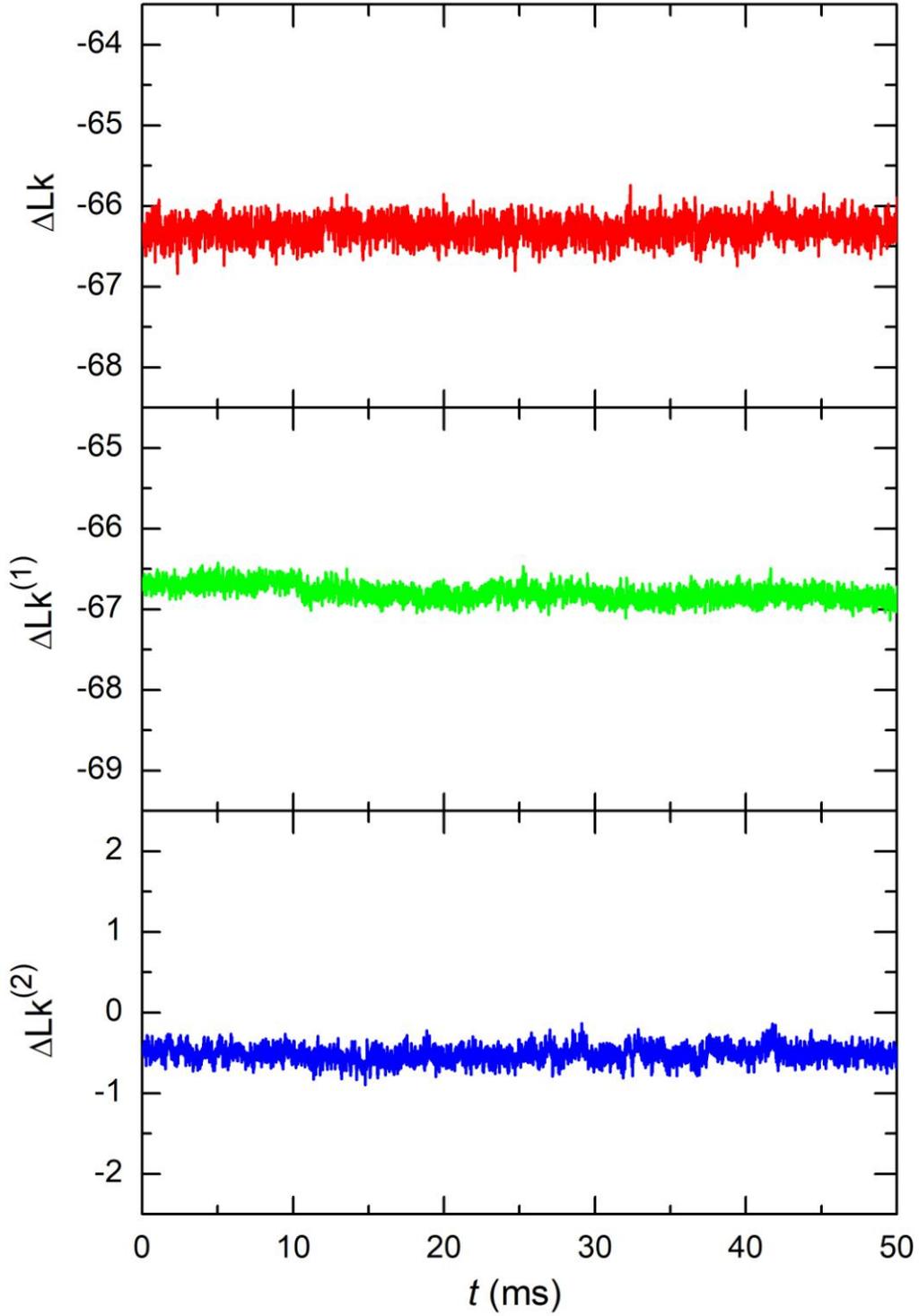

**Figure S1 :** Time evolution of the total linking number difference $\Delta L\mathrm{k}$ **(top)**, as well as the linking number difference in each loop, $\Delta L\mathrm{k}^{(1)}$ **(middle)** and $\Delta L\mathrm{k}^{(2)}$ **(bottom)**, for a simulation performed with a Model II topological barrier that separates the DNA circular chain into two equal loops with respective helical density $\sigma^{(1)} = -0.063$ and $\sigma^{(2)} = 0$. Parameters of the simulation were $\alpha = n/2 = 1440$, $\beta = n = 2880$, $\gamma = 3n/4 = 2160$, and $\Omega = 0.9 \text{ rad } \mu\text{s}^{-1}$.

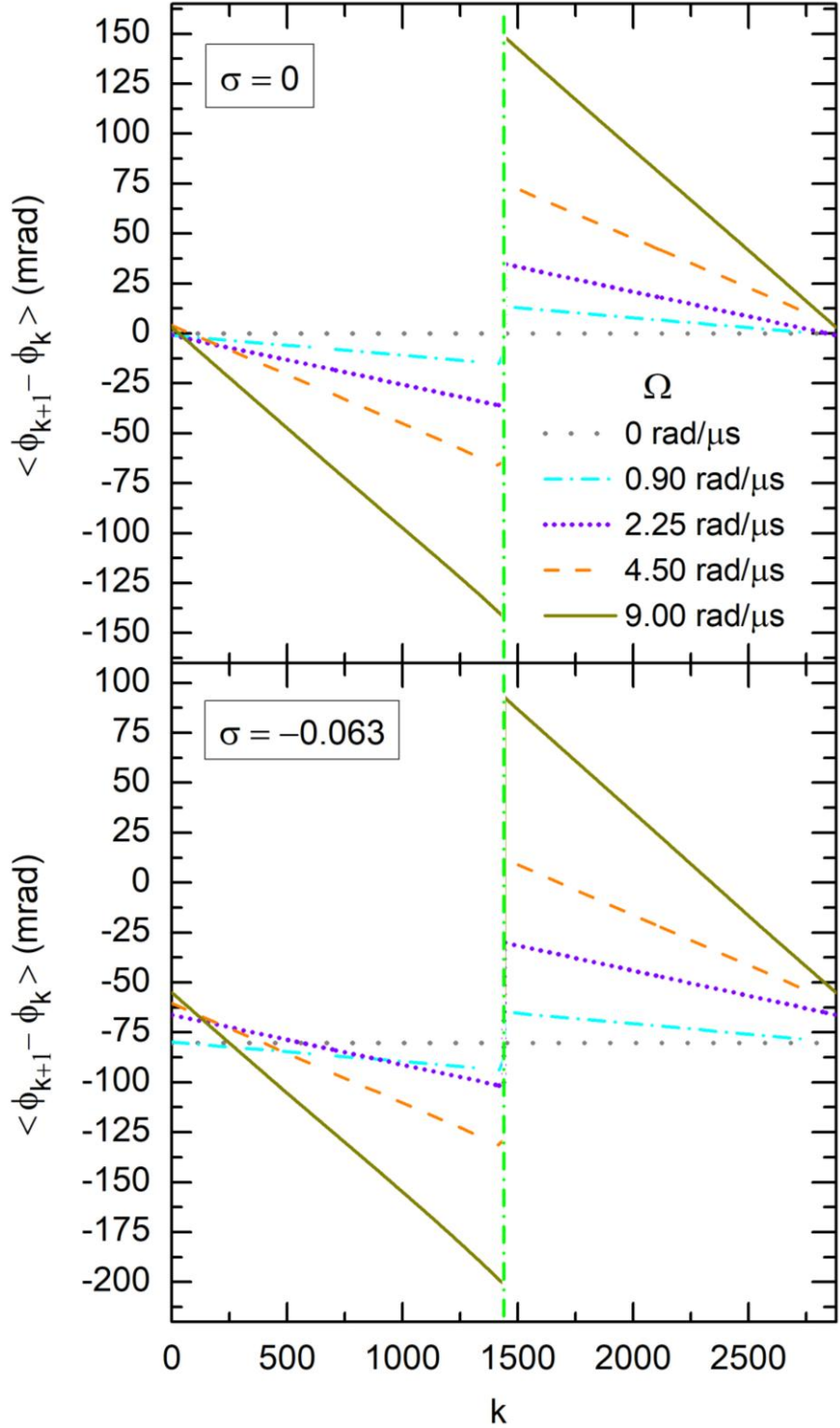

**Figure S2.** Evolution of $<\Phi_{k+1} - \Phi_k>$, the time average of the variation of the torsional angle between beads $k$ and $k+1$, as a function of DNA bead index $k$, in the absence of topological barriers and RNAP translation, for a DNA chain with $\sigma=0$ **(top)** or $\sigma=-0.063$ **(bottom)**. Simulations were performed with $\gamma = n/2 = 1440$ and 4 values of $\Omega$ ranging from 0.9 to 9.0 rad μs$^{-1}$.

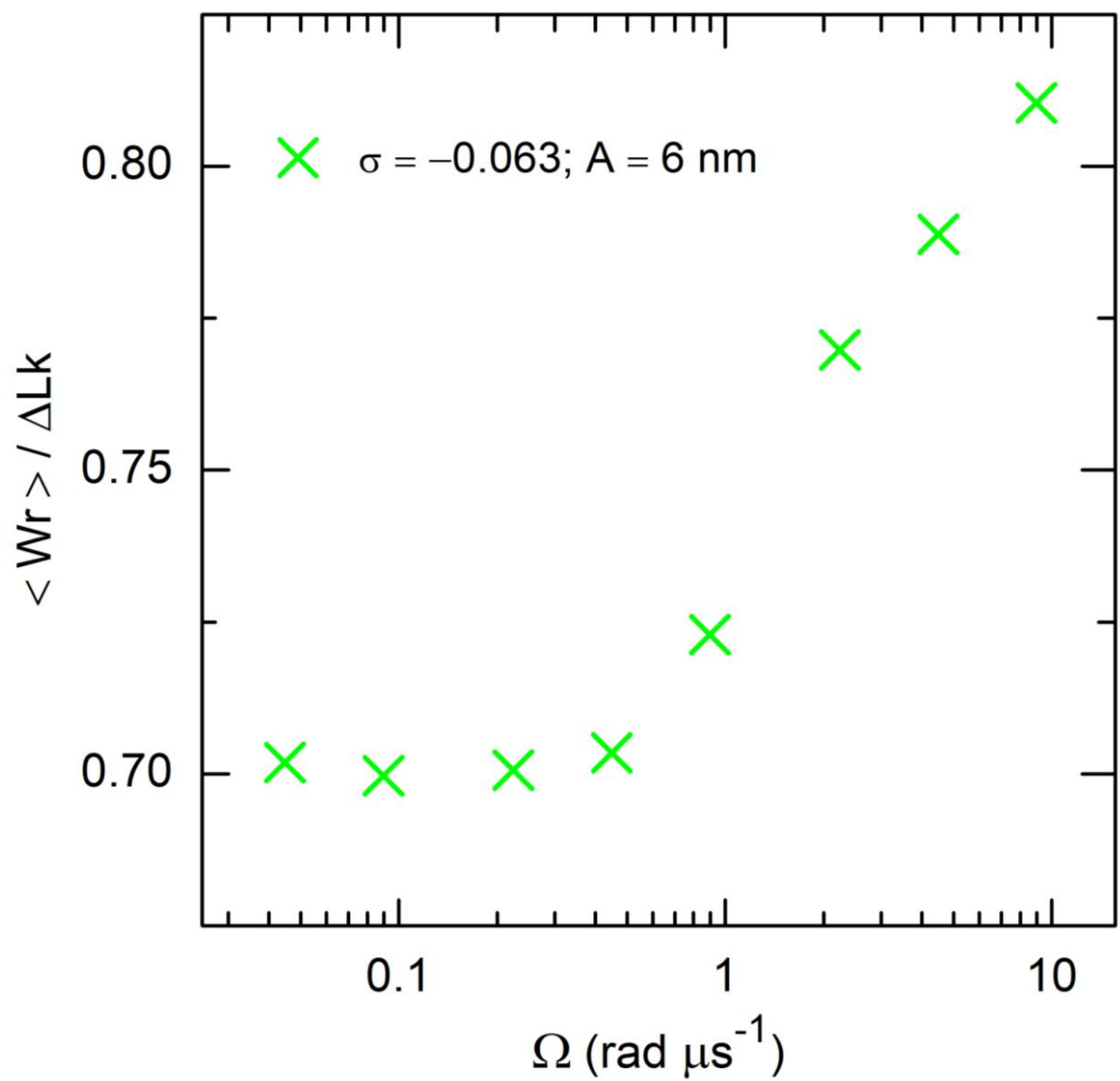

**Figure S3.** Evolution of $\langle Wr \rangle / \Delta Lk$, the contribution of the mean writhe to the linking number of the DNA chain, as a function of $\Omega$, the RNAP active rotational speed, in the absence of topological barriers and RNAP translation. Simulations were performed with $\sigma = -0.063$ and $\gamma = n/2 = 1440$.

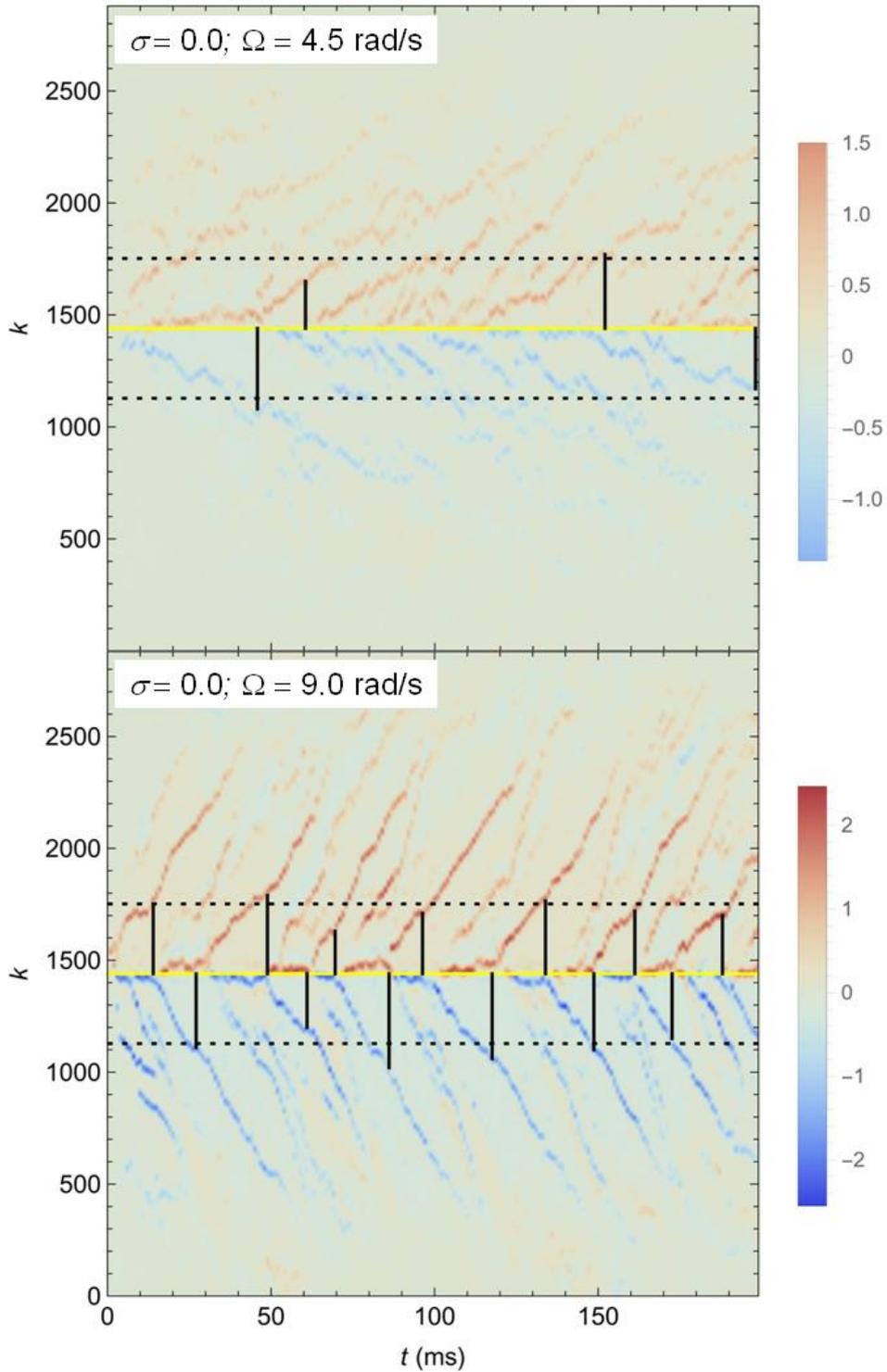

**Figure S4.** Density plots showing the time evolution of the local writhe $Wr(k, l_w)$ for all DNA beads $k$, computed from simulations with $\sigma = 0$, $\gamma = n/2 = 1440$ and $\Omega = 4.5$ rad µs$^{-1}$ **(top)** or $\Omega = 9.0$ rad µs$^{-1}$ **(bottom)**, in the absence of topological barriers and RNAP translation. The horizontal yellow lines indicate the position of the RNAP. Vertical black segments indicate those moments when a long plectoneme detaches from one side of the RNAP, while another plectoneme starts growing on the other side. Horizontal black dashed lines indicate the average half-length at which plectonemes detach from the RNAP.

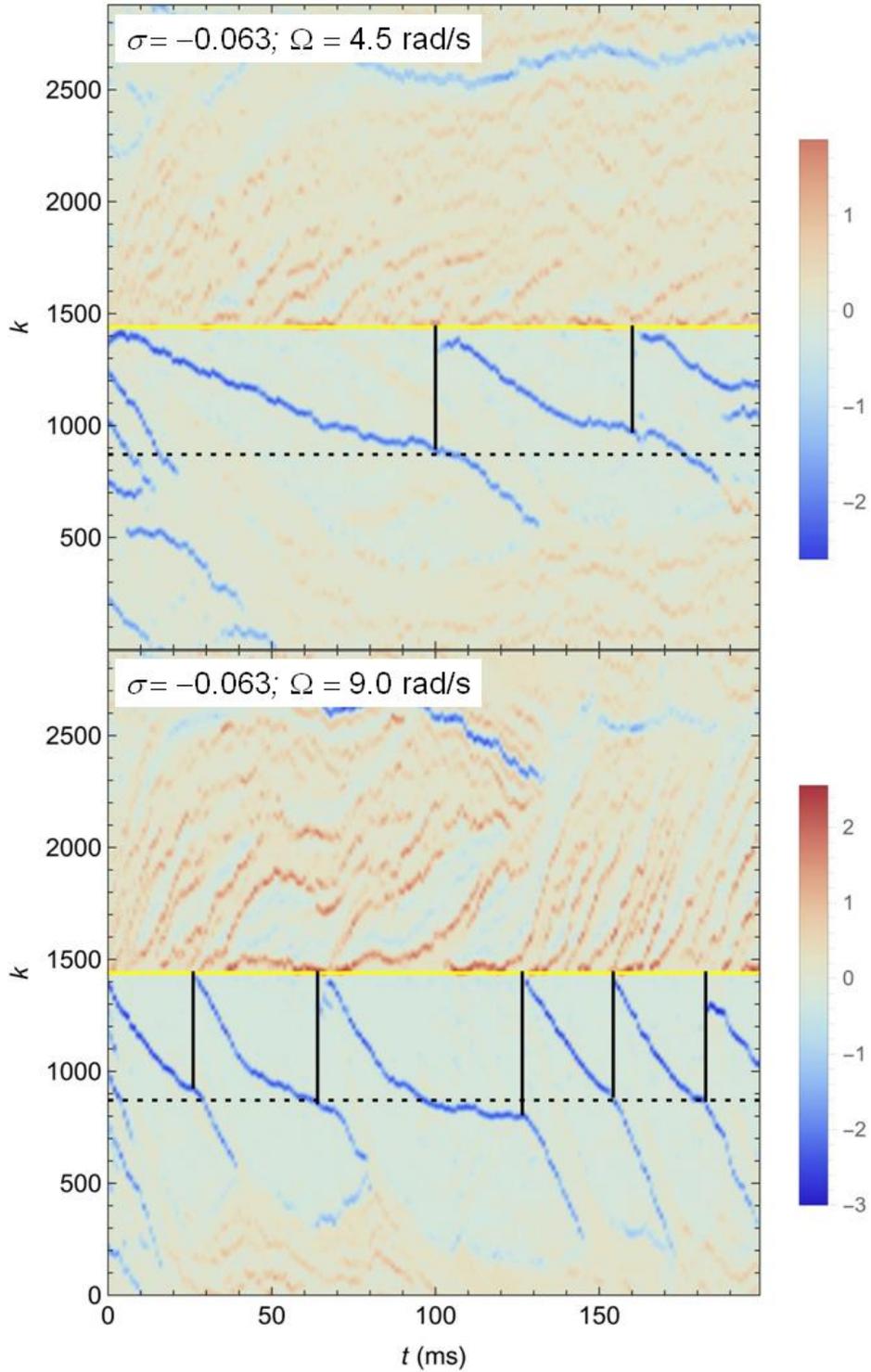

**Figure S5.** Density plots showing the time evolution of the local writhe $Wr(k, l_w)$ for all DNA beads $k$, computed from simulations with $\sigma = -0.063$, $\gamma = n/2 = 1440$ and $\Omega = 4.5$ rad µs$^{-1}$ **(top)** or $\Omega = 9.0$ rad µs$^{-1}$ **(bottom)**, in the absence of topological barriers and RNAP translation. The horizontal yellow lines indicate the position of the RNAP. Vertical black segments indicate those moments when a long negatively supercoiled plectoneme detaches from the upstream side of the RNAP. Horizontal black dashed lines indicate the average half-length at which plectonemes detach from the RNAP.

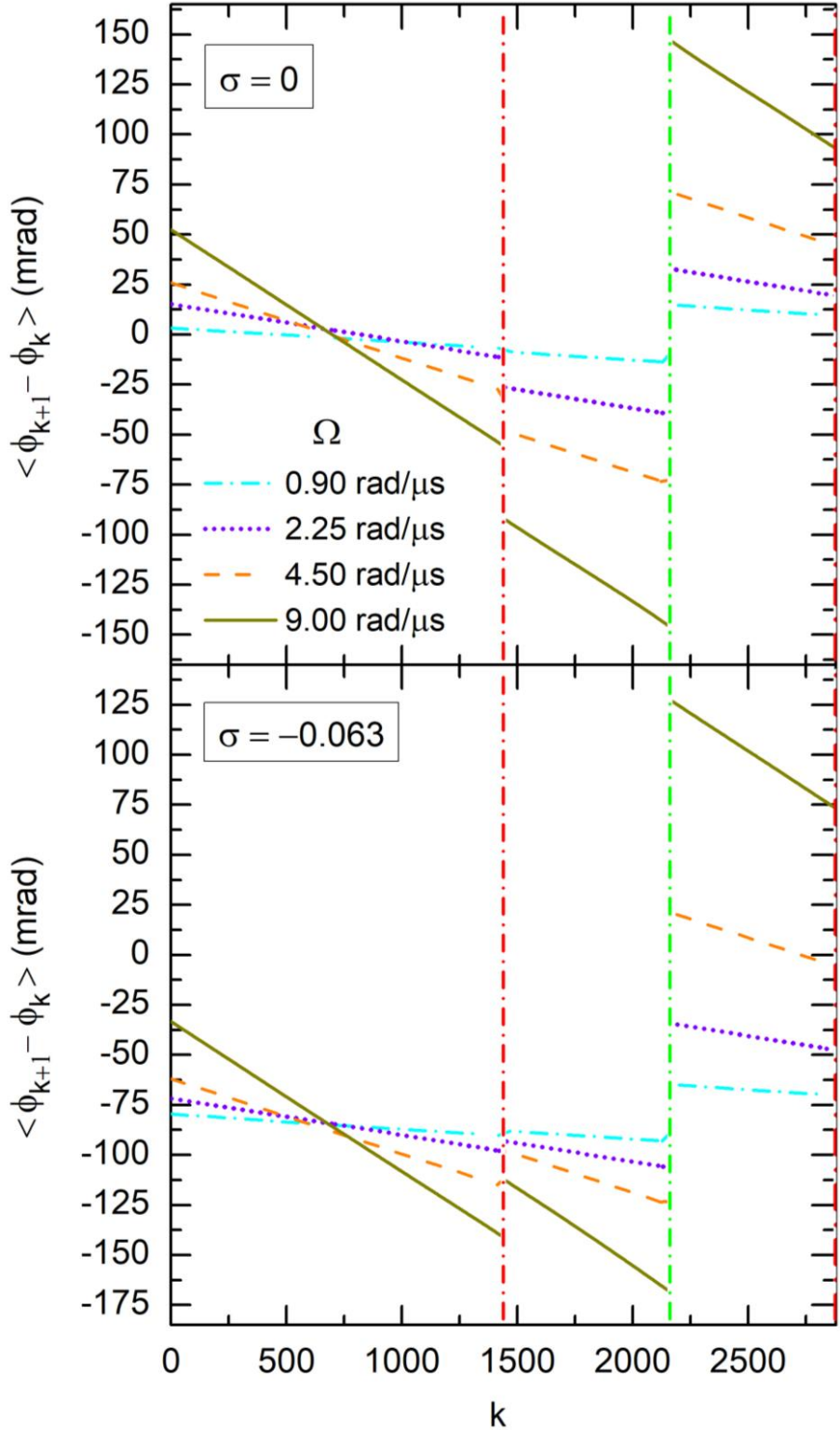

**Figure S6.** Evolution of $<\Phi_{k+1} - \Phi_k>$, the time average of the variation of the torsional angle between beads $k$ and $k+1$, as a function of DNA bead index $k$, in the presence of a Model I topological barrier but in the absence of RNAP translation, for a DNA chain with $\sigma=0$ **(top)** or $\sigma=-0.063$ **(bottom)**. Simulations were performed with $\alpha = n/2 = 1440$, $\beta = n = 2880$, $\gamma = 3n/4 = 2160$ and 4 values of $\Omega$ ranging from 0.9 to 9.0 rad µs$^{-1}$.

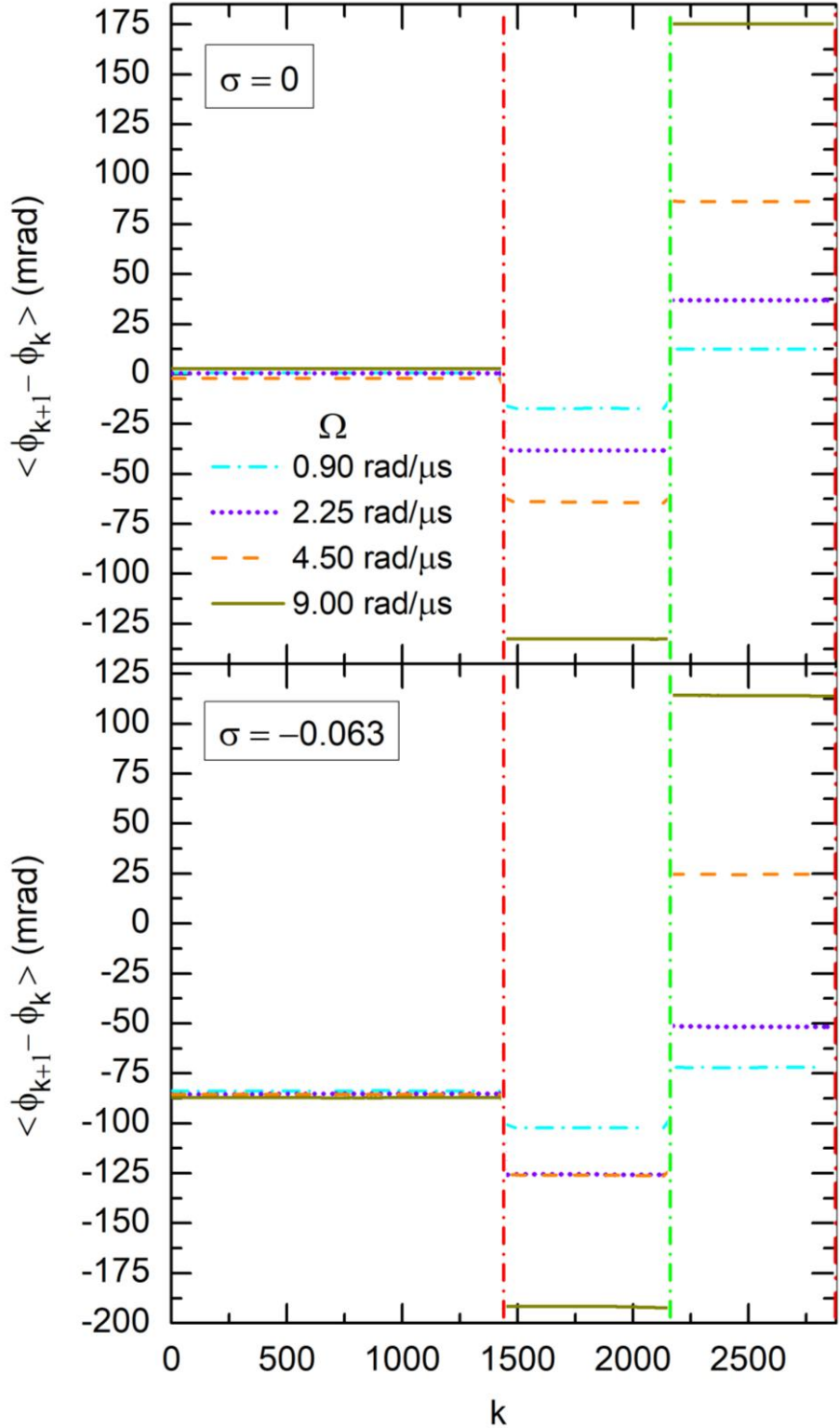

**Figure S7.** Evolution of $<\Phi_{k+1} - \Phi_k>$, the time average of the variation of the torsional angle between beads $k$ and $k+1$, as a function of DNA bead index $k$, in the presence of a Model II topological barrier but in the absence of RNAP translation, for a DNA chain with $\sigma = 0$ **(top)** or $\sigma = -0.063$ **(bottom)**. Simulations were performed with $\alpha = n/2 = 1440$, $\beta = n = 2880$, $\gamma = 3n/4 = 2160$ and 4 values of $\Omega$ ranging from 0.9 to 9.0 rad µs$^{-1}$.

## MOVIE CAPTIONS

**Movie S1.** AVI file. 2000 frames shown at 20 fps and encompassing a real time window of 200 ms. Sequence of contact maps obtained in the course of a simulation with $\sigma=0$, $\gamma = n/2 = 1440$ and $\Omega = 9.0$ rad $\mu s^{-1}$, in the absence of topological barriers and RNAP translation. The horizontal and vertical green dashed lines indicate the position of the RNAP.

**Movie S2.** AVI file. 1000 frames shown at 20 fps and encompassing a real time window of 100 ms. Sequence of contact maps obtained in the course of a simulation with $\sigma=0$ and $\Omega = 9.0$ rad $\mu s^{-1}$, with RNAP translation but in the absence of topological barriers. The horizontal and vertical green dashed lines indicate the position of the RNAP.

**Movie S3.** AVI file. 1600 frames shown at 20 fps and encompassing a real time window of 160 ms. Sequence of contact maps obtained in the course of a simulation with $\sigma = -0.063$, $\gamma = n/2 = 1440$ and $\Omega = 0.9$ rad $\mu s^{-1}$, in the absence of topological barriers and RNAP translation. The horizontal and vertical green dashed lines indicate the position of the RNAP.

**Movie S4.** AVI file. 2000 frames shown at 20 fps and encompassing a real time window of 200 ms. Sequence of contact maps obtained in the course of a simulation with $\sigma = -0.063$, $\gamma = n/2 = 1440$ and $\Omega = 9.0$ rad $\mu s^{-1}$, in the absence of topological barriers and RNAP translation. The horizontal and vertical green dashed lines indicate the position of the RNAP.

**Movie S5.** AVI file. 800 frames shown at 20 fps and encompassing a real time window of 80 ms. Sequence of contact maps obtained in the course of a simulation with $\sigma = -0.063$ and $\Omega = 9.0$ rad $\mu s^{-1}$, with RNAP translation but in the absence of topological barriers. The horizontal and vertical green dashed lines indicate the position of the RNAP.

**Movie S6.** AVI file. 1000 frames shown at 20 fps and encompassing a real time window of 100 ms. Sequence of contact maps obtained in the course of a simulation with $\sigma = 0$, $\gamma = 3n/4 = 2160$ and $\Omega = 9.0$ rad µs$^{-1}$, in the presence of a Model II topological barrier bridging beads $\alpha = n/2 = 1440$ and $\beta = n = 2880$, but in the absence of RNAP translation. The horizontal and vertical green dashed lines indicate the position of the RNAP, the red ones indicate the positions of the beads bridged by the topological barrier.